\documentclass[12pt,a4paper]{article}

\usepackage{graphicx}
\usepackage{color}
\usepackage{natbib}
\usepackage{amsmath}
\usepackage{eurosym}
\usepackage{url}
\usepackage{fancyhdr}
\usepackage{tikz}
\usepackage{rotating}

\usepackage{multirow}%
\usepackage{amssymb,amsfonts}%
\usepackage{amsthm}%
\usepackage{mathrsfs}%
\usepackage[title]{appendix}%
\usepackage{xcolor}%
\usepackage{textcomp}%
\usepackage{manyfoot}%
\usepackage{booktabs}%
\usepackage{algorithm}%
\usepackage{algorithmicx}%
\usepackage{algpseudocode}%
\usepackage{listings}%

\usepackage{bm}

\newcommand{\by}{\mathbf{y}}

\newcommand{\ba}{\mathbf{a}}

\newcommand{\bS}{\mathbf{S}}

\newcommand{\bI}{\mathbf{I}}

\newcommand{\btheta}{\boldsymbol{\theta}}

\newcommand{\boldeta}{\boldsymbol{\eta}}
\newcommand{\bmu}{\boldsymbol{\mu}}

\newcommand{\cN}{\mathcal{N}}
\newcommand{\cG}{\mathcal{G}}

\newcommand{\cC}{\mathcal{C}}

\newcommand{\ym}{{\mathbf y}} 

\newcommand{\etav}{{\boldsymbol{\eta}}} 
\newcommand{\Gamfun}[1]{\Gamma (#1)}

\newcommand{\Prob}[1]{\mbox{\rm Pr}\{#1\}}

\newcommand{\EE}{\mbox{\rm E}}
\newcommand{\Ew}[1]{\EE(#1)}   
\newcommand{\VV}{\mbox{\rm V}}
\newcommand{\V}[1]{\VV (#1)}

\newcommand{\pkg}[1]{{\fontseries{m}\fontseries{b}\selectfont #1}}

\newcommand{\cb}{c_{\phi}}
\newcommand{\db}{d_{\phi}}

\begin{document}
	
\title{Without Pain -- Clustering Categorical Data Using a Bayesian Mixture of Finite Mixtures of Latent Class Analysis Models}
	
\author{Gertraud Malsiner-Walli \thanks{WU Vienna University of Economics and Business}, Bettina Gr\"un
		\thanks{WU Vienna University of Economics and Business}\\
		and Sylvia Fr\"uhwirth-Schnatter \thanks{WU Vienna University of Economics and Business}
		\vspace{0.5cm}}

\date{}

\maketitle

{\small We propose a Bayesian approach for model-based clustering
	of multivariate categorical data where variables are allowed to be
	associated within clusters and the number of clusters is
	unknown. The approach uses a two-layer  mixture of finite mixtures
	model where the cluster distributions are approximated using latent
	class analysis models.  A careful specification of priors with
	suitable hyperparameter values is crucial to identify the two-layer
	structure and obtain a parsimonious cluster solution.  We outline
	the Bayesian estimation based on Markov chain Monte Carlo sampling
	with the telescoping sampler and describe how to obtain an
	identified clustering model by resolving the label switching issue.
	Empirical demonstrations in a simulation study using artificial data
	as well as a data set on low back pain indicate the good clustering
	performance of the proposed approach 
	provided hyperparameters are	selected that
	induce sufficient shrinkage. }

\newpage
\section{Introduction}\label{sec:introduction}

	In this paper, we develop a model-based clustering approach for
	multivariate categorical data. All variables are assumed
	cluster-relevant but they are potentially associated within
	clusters. The approach also infers a suitable number of clusters. 
In particular, the motivation for developing this approach is a
specific data set investigating low back pain disorders described in
\citet{lca:FopetAl2017}.  A questionnaire was developed for patients
by experienced physiotherapists where the answers allow to classify
patients according to the type of low back pain they are suffering
from. The questionnaire consists of a range of binary questions about
the presence or absence of certain pain symptoms.  As the list of
clinical criteria was assembled by experts, all variables are expected
to possess good discriminatory power and, hence, are potentially
cluster-informative.  However, it is also suspected that some of the
indicators carry the same information about the low back pain type,
i.e., they are redundant.  Additionally, variables are expected to be
dependent within the low back pain types due to the same underlying
neurophysiological mechanisms being responsible for the pain
generation and the manifestation captured by several binary
indicators.  To address 
	this association within clusters, we
	propose 
an extension of the latent class analysis model for
	Bayesian model-based clustering where all variables (also the ``redundant''
	ones) are included in the analysis and where within-group associations
	between variables are accounted for.
	
	Latent class analysis \citep[LCA;][]{Lazarsfeld:1950, Goodman:1974} is
	a well-established framework which has been developed to model
	dependence among multivariate categorical variables. LCA assumes that
	the dependency structure between categorical variables observed in the
	data is due to the presence of latent groups called classes and that
	conditional on class membership the categorical variables are
	independent.  This model class thus represents a convenient way to
	capture dependencies between categorical variables.  However, assuming
	conditionally independent variables, i.e., that classes are fully
	characterized by the occurrence probability of each category of a
	variable, is very restrictive in a clustering context.  The sparsity
	of the model, implied by the conditionally independence assumption, is
	of great advantage, but in 
		clustering applications this
		assumption might be questionable.  In the presence of within-cluster
		associations, the standard LCA model will fit more classes than there
		are clusters in the data, in order to provide an adequate fit to the
		data.  As a consequence, the assumption of a one-to-one relationship
		between the classes of the LCA model and the clusters in the data is
		no longer applicable, degrading the interpretability of the estimated
		clusters.
		
		In the literature, approaches to account for association between
		categorical variables within clusters follow basically two streams of
		research.  The first approach consists in selecting a subset of the
		observed variables in order to extract independent, cluster-relevant
		variables, 
		see, e.g., \citet{lca:Maugis2009,
			lca:DeanRaftery2010, 
			lca:BartolucciMontanariPandolfi2016,
			lca:WhiteWyseMurphy2016,
			lca:FopetAl2017}.  The second approach consists
		in modeling explicitly the conditional dependencies in the classes.
		In log-linear models \citep{lca:Bock1986, lca:Agresti2002,
			lca:PapathomasRichardson2016}, the frequencies in the contingency
		table obtained by cross-tabulating the categorical variables are
		modeled by including suitable interaction terms between
		variables. Multilevel latent class models \citep{lca:Vermunt2003}
		assume that the cluster-specific distribution of the categorical
		variables depends on continuous latent variables which allow to model
		the dependencies among the observed categorical variables in the
		latent space.  \citet{lca:GolliniMurphy2014} propose a mixture of
		latent trait analyzers to accommodate dependencies within a cluster.
		\citet{lca:Marbac2015} investigate a block extension of the LCA model,
		where conditional on the class, the variables are grouped into
		independent blocks, with each block following a specific distribution
		which takes into account the dependency between variables.
		
		In this work, we combine the original idea of the LCA model, namely to
		capture associated categorical variables by the introduction of a
		latent group variable, with a model-based clustering approach. We
		propose a mixture of LCA models which results in a two-layer model. On
		the upper level, the mixture approach is used for model-based
		clustering of the data, while on the lower level the
		component-specific LCA model represents a cluster in the data and is
		used to approximate the cluster distribution, allowing for association
		between variables within a cluster to be captured.  
			On the lower
			level, each cluster is assumed to be composed of several classes to
			capture the association within variables and these classes thus do
			not correspond to identifiable groups. 
		In the following we will
		call the (possibly associated) components of the mixture distribution on
		the upper level ``components'' or ``clusters'', while the classes of
		the LCA models on the lower level are called ``subcomponents'' or
		``classes''.
		
		Statistical inference of finite mixtures is generally complicated due
		to spurious modes and unboundedness of the mixture likelihood (see,
		e.g., \citealt{Mix:Fruehwirth2006}, chap.~2). But the estimation of a
		two-layer mixture model is particularly challenging due to its
		identifiability issues.  Exchanging subcomponents on the lower level
		between components of the upper level leads to different
		component-specific distributions, but overall the mixture density
		remains the same. Thus, a two-layer mixture model is not identifiable
		based on the mixture likelihood alone in the absence of additional
		information. This is especially 
			problematic 
		in the context of clustering,
		as only identified clusters are interpretable, see a recent discussion
		in \cite{lca:GuDunson2023} in the framework of deep learning models.
		
		In the Bayesian setting, we resolve the identifiability issue of the
		two-layer mixture model by specifying suitable priors which favor a
		unique assignment of the classes in the LCA models on the lower level
		to components on the upper level.  For a mixture of Gaussian mixtures,
		i.e., a two-layer mixture model for multivariate continuous data,
		\citet{Malsiner-Walli+Fruehwirth-Schnatter+Gruen:2017} suggested a
		prior which induces shrinkage of the subcomponent means on the lower
		level toward 
			a cluster-specific center on the upper level
		to obtain
		identifiability.  In this paper, we investigate how the approach
		developed for mixtures of multivariate Gaussian mixtures can be
		extended and adapted to mixtures of LCA models. We propose a prior
		specification which induces shrinkage of the subcomponent occurrence
		probabilities toward a central, cluster-specific occurrence
		probability vector.  This is related to
		\cite{lca:DuranteDunsonVogelstein2017} who, in the framework of
		network-valued data, define the component-specific edge-probability
		vector as a function of a shared similarity vector, common to all
		components, and a component-specific deviation vector, which is pulled
		toward zero by a shrinkage prior. However, in contrast to
		\cite{lca:DuranteDunsonVogelstein2017}, who operate in the underlying
		latent continuous space, we model the common ``center'' of the
		occurrence probabilities of the classes belonging to the same cluster
		and also their ``precision'' around the common center directly,
		without  referring  to any 
		latent space representation.
		
		Our proposed Bayesian mixture of LCA models approach also addresses
		the issue of selecting the number of clusters in the data. If the
		number of clusters in the data is unknown, in the Bayesian setting a
		prior on the number of components in the mixture distribution can be
		specified, and the resulting model is referred to as a ``mixture of
		finite mixtures (MFM) model''
		\citep{FM:MillerHarrison2018,Fruehwirth-Schnatter+Gruen+Malsiner-Walli:2021}.
		For a MFM model, inference is challenging as a transdimensional
		sampler is required which is able to jump between parameter spaces of
		varying dimension.  We 
			use a Gibbs sampling scheme with data
			augmentation for Markov chain Monte Carlo (MCMC) inference which
			relies on the telescoping sampler proposed by
			\citet{Fruehwirth-Schnatter+Gruen+Malsiner-Walli:2021} as an
			alternative to other, more complex approaches such as Reversible Jump
			MCMC \citep{Mix:RichardsonGreen1997} or Chinese Restaurant process based
			sampling schemes popular in Bayesian non-parametrics
			\citep{FM:MillerHarrison2018}. An advantage of the telescoping sampler
			is that sampling of the number of components is performed conditional
			on the current partition of the data into non-empty clusters and
			completely independent of the component distributions. This makes the
			sampler easy to implement when fitting a Bayesian mixture of LCA
			models where the components are part of a multi-layer model.
			
			The rest of the article is organized as
			follows. Section~\ref{sec:mixture-latent-class} provides the model
			specification including a detailed discussion of the selected priors
			to obtain a Bayesian mixture of LCA models suitable for clustering
			multivariate categorical data. Model estimation and identification are
			discussed in Section~\ref{sec:posterior-inference}. The performance of
			the proposed strategy is evaluated in Section~\ref{sec:empir-demonstr}
			in a simulation study and for the low back pain data set already used
			in \citet{lca:FopetAl2017}. Section~\ref{sec:summary} concludes.
			
			\section{Bayesian mixture of finite mixtures of latent class analysis
				models}\label{sec:mixture-latent-class}
			
			\subsection{Data model}\label{sec:data-model}
			We consider multivariate categorical observations $\bm{y}_i$,
			$i=1,\ldots,N$, consisting of $r$ variables where each variable
			$j=1,\ldots,r$ is categorical with $D_j$ categories. Each observation
			is given by $\bm{y}_i = (y_{ij})_{j=1,\ldots,r}$ with
			$y_{ij} \in \{1,\ldots,D_j\}$.  The $N$ observations $\bm{y}_i$ are
			assumed to be generated independently from a finite mixture
			distribution with $K$ components with density
			\begin{align}\label{eq:comp}
				p(\bm{y}_i | K, \bm{\Theta}_K, \bm{\eta}_K) &= \sum_{k=1}^{K} \eta_k p_k(\bm{y}_i | \bm{\theta}_k),
			\end{align}
			where, for a fixed number of components $K$,
			$\bm{\Theta}_K=(\bm{\theta}_k)_{k=1,\ldots,K}$ is the vector of
			component-specific parameters and
			$\bm{\eta}_K = (\eta_k)_{k=1,\ldots,K}$ are the positive component
			weights with
			\begin{align*}
				\sum_{k=1}^{K} \eta_k & = 1.
			\end{align*}
			Each component distribution $p_k(\bm{y}_i | \bm{\theta}_k)$ is given
			by a LCA model, i.e.,
			\begin{align}
				p_k(\bm{y}_i | \bm{\theta}_k) &= \sum_{l=1}^{L} w_{kl} \prod_{j=1}^r\prod_{d=1}^{D_j} \pi_{kl,jd}^{I\{y_{ij} = d\}},\label{eq:subcomp}
			\end{align}
			where $L$ is the number of classes (subcomponents) of component $k$
			and $\bm{w}_{k} = (w_{kl})_{l=1,\ldots,L}$ are the positive
			subcomponent weights with
			\begin{align*}
				\sum_{l=1}^{L} 
				w_{kl} & = 1.
			\end{align*}
			Note that we assume that the number of classes (subcomponents) $L$
				is the same across all components. In principle one could also
				specify a value $L_k$ which varies across components and is learned from the data. 

			The component-specific parameter $\bm{\theta}_k$ consists of
			$(\bm{w}_k, \bm{\pi}_k)$, where
			$\bm{\pi}_k = (\pi_{kl,jd})_{d=1,\ldots,D_j; j=1,\ldots, r;
				l=1,\ldots, L }$ with $\pi_{kl,jd}$ equal to the occurrence
			probability of observation $y_{ij}$ taking category $d$ conditional on
			being from subcomponent $l$ of component $k$.  The vector of
			subcomponent weights is given by
			$\bm{w}_{KL} = (\bm{w}_{k})_{k=1,\ldots,K}$ and the vector of
			subcomponent occurrence probabilities by
			$\bm{\pi}_{KL} = (\bm{\theta}_k)_{k=1,\ldots,K}$.
			
			\subsection{Prior specification}\label{sec:prior-specification}
			
			The data model specified in \eqref{eq:comp} and \eqref{eq:subcomp} is
			complemented by priors for the model parameters
			$(K, \bm{\eta}_K, L, \bm{w}_{KL}, \bm{\pi}_{KL})$. 
				In the
				following, we carefully select priors to achieve the following
				modeling aims: (1) estimation of a parsimonious number of clusters
				in the data and (2) identification of the cluster-specific distributions.  We
				use conditionally conjugate priors, where possible, to ease
				inference. Further, we impose exchangeable priors across the
				components and subcomponents to preserve prior invariance and
				symmetry with respect to arranging the components in a mixture
				distribution. 
			
			\subsubsection{Determining the number of components and
				subcomponents}\label{sec:SpecKL}
			
			In clustering applications, the number of components $K$ in the upper
			level model \eqref{eq:comp} is usually not known.  In the Bayesian
			framework, it is natural to treat $K$ as a random variable and to
			specify a prior on it, i.e.,
			\begin{align*}
				K&\sim p(K).
			\end{align*}
			This results in a mixture
			of finite mixtures model \citep[MFM;][]{FM:MillerHarrison2018} and
			posterior inference for $K$ can be employed to
			determine a suitable number of components. 
			
			\citet{Fruehwirth-Schnatter+Gruen+Malsiner-Walli:2021} review previous
			suggestions for the prior on $K$ and propose to use the shifted
			beta-negative-binomial (BNB) distribution for $K$. This distribution
			encompasses other distributions previously proposed as special cases
			(i.e., the Poisson, the geometric and the negative-binomial
			distribution). The shifted BNB distribution is a three-parameter
			distribution which allows to not only to calibrate the prior mean and
			variance, but also to control separately the probability of the
			homogeneity model $K = 1$ and the tail behavior, i.e., the mass
			assigned to large values of $K$. Following
			\citet{Fruehwirth-Schnatter+Gruen+Malsiner-Walli:2021} and
			\citet{lca:Gruen+Malsiner-Walli+Fruehwirth-Schnatter:2022}, we propose
			to use $K - 1 \sim \mathit{BNB}(1,4,3)$ as a suitable prior on $K$ in
			a clustering application. This prior concentrates a lot of mass on the
			homogeneity model and assigns decreasing prior weights to increasing
			number of components, while still having some mass assigned to larger
			values.  See Appendix~\ref{sec:beta-negat-binom} for more details on
			the BNB distribution.
			
			On the lower level, also the number of classes 
				is typically
				unknown for each component $k$. However, on this level we are not
				interested in estimating the ``true'' number of classes 
				forming
				cluster $k$, as there is no clustering task on this level. We only
				need to ensure that this number 
			is chosen sufficiently large to be flexible
			enough to capture the association structure in cluster $k$.  
				Similar to kernel density estimation, where the exact number of
				kernels is usually not of interest as long as the approximation is
				sufficiently close, we select a fixed, sufficiently large value $L$
				for the number of classes across all components.  However, a too
				generous choice of $L$ should be avoided for computational reasons. 
			In our analyses we set $L$ to a fixed value varying between $2$ and
			$5$ 
				which is sufficient to capture the within-cluster associations
				in our applications.
			
			The priors on the component weights $\bm{\eta}_K$ on the upper level
			as well as the subcomponent weights $\bm{w}_k$, $k=1,\ldots,K$, on the
			lower level are specified by symmetric Dirichlet distributions with
			parameters $\gamma_K$ and $\delta$, respectively:
			\begin{align*}
				\bm{\eta}_K|K &\sim \mathcal{D}_K(\gamma_K), &
				\bm{w}_k &\sim \mathcal{D}_{L}(\delta),\quad  k=1,\ldots,K,
			\end{align*}
			where $\mathcal{D}_{p}(a)$ is the symmetric $p$-dimensional Dirichlet
			distribution with scalar parameter $a$.
			
			For the Dirichlet parameter $\gamma_K$ on the upper level, we use a
			specification which depends on the number of components $K$.  This
			choice is based on
			\cite{Fruehwirth-Schnatter+Gruen+Malsiner-Walli:2021} who emphasize
			the crucial distinction between $K$, the number of components in the
			mixture distribution, and $K_+$, the number of ``filled'' components,
			i.e., the number of components which generated the data at hand and to
			which observations are assigned during MCMC sampling with data
			augmentation. Only filled components correspond to clusters in the
			data. When estimating the number of clusters in the data, the
			posterior of $K_+$ is of crucial interest whereas the posterior of $K$
			rather reflects the potential number of clusters in the population and
			not in the sample \citep{FM:McCullagh2008many}.
			\citet{Fruehwirth-Schnatter+Gruen+Malsiner-Walli:2021} and
			\citet{lca:Greve+Gruen+Malsiner-Walli:2022} show that the prior on the
			number of data clusters $K_+$ is crucially influenced by the choice of
			$\gamma_K$, the Dirichlet parameter of the prior on the weight
			distribution. The smaller $\gamma_K$, the less components $K_+$ are a
			priori assumed to have generated the data at hand.
			
			In particular, we follow
			\citet{Fruehwirth-Schnatter+Gruen+Malsiner-Walli:2021} and consider
			the dynamic specification of a MFM model with $\gamma_K =
			\alpha/K$. In this way, the Dirichlet parameter is not a fixed value
			but decreases with increasing number of components $K$, implying an
			increasing gap between $K$ and $K_+$ for larger values of $K$. This
			dynamic Dirichlet prior for the component weights behaves like a
			shrinkage prior as it favors smaller values of $K_+$ when $K$ is
			increasing.  Additionally, in order to learn from the data the
			appropriate value of $\alpha$ for the specification of
			$\gamma_K=\alpha/K$, we specify a gamma hyperprior on $\alpha$, i.e.,
			$\alpha \sim \cG(a,b)$ with expectation $\Ew{\alpha}=a/b$. In our
			empirical demonstrations, we use $\alpha \sim \cG(1,2)$ 
				to induce  sparsity. 
			
			In contrast, when selecting $\delta$ for the Dirichlet prior on the
			class weights on the lower level, no shrinkage of the number of
			``filled'' classes to a value smaller than $L$ is required. We would
			like to be as flexible as possible in regard to the class size
			distribution among the classes and specify $\delta=1$, which
			corresponds to the uniform distribution on the unit simplex. 
			
			\subsubsection{Identifying the cluster distributions}\label{sec:hierPrior}
			
			As described in Section~\ref{sec:introduction}, the likelihood of the
			two-layer mixture of LCA models is invariant to how the latent classes
			are assigned to the different components.  In the Bayesian setting, it
			is possible to achieve model identification by specifying suitable
			prior distributions which favor a unique assignment of subcomponents
			to clusters in the posterior distribution.  For mixtures of
			multivariate Gaussians,
			\citet{Malsiner-Walli+Fruehwirth-Schnatter+Gruen:2017} proposed to use
			priors which induce shrinkage of the subcomponent means of the same
			component towards a common cluster-specific mean.  Generalizing this
			approach to the mixture of LCA models approach leads to the following
			hierarchical prior construction.
			
			On the lower level, for each class $l$ of the LCA model defining the $k$th cluster, the conditionally conjugate
			prior for the class occurrence probabilities $\bm{\pi}_{kl,j}$ of
			variable $j$ is specified by a Dirichlet distribution, i.e.,
			\begin{align*}
				\bm{\pi}_{kl,j} &\sim \mathcal{D}_{D_j}(\bm{\alpha}_{k,j}), \quad l=1,\ldots,L,  \quad k=1,\ldots,K,  \quad j=1,\ldots,r,
			\end{align*}
			where $\mathcal{D}_{p}(\ba)$ is the $p$-dimensional Dirichlet
			distribution with parameter vector $\ba$. 
			Note that this prior is exchangeable across the $L$ classes for each combination of $k$ and $j$.
			We decompose the Dirichlet
			parameter vector $ \bm{\alpha}_{k,j}$ into the location parameter
			$\bmu_{k,j}=(\mu_{k,jd})_{d=1,\ldots,D_j}$
			which defines an ``average'' probability distribution
			over the $D_j$ categories of variable $j$ in cluster $k$ and the
			scalar precision parameter $\phi_{k,j}$ taking values in $\Re^+$ which control variation around this average:
			\begin{align*}
				\bm{\alpha}_{k,j} &= \bm{\mu}_{k,j} \phi_{k,j}. 
			\end{align*}
				Note that this decomposition leads to the expected occurrence probability   and its variance  given by 
				\begin{equation*}
					\Ew{\pi_{kl,jd}} = \frac{\mu_{k,jd}}{\sum_{d=1}^{D_j} \mu_{k,jd}}, \quad
					\V{\pi_{kl,jd}} = \frac{ \Ew{\pi_{kl,jd}} (1-\Ew{\pi_{kl,jd}})}{\phi_{k,j}\sum_{d=1}^{D_j} \mu_{k,jd} + 1},
					\quad l=1, \ldots,L.
				\end{equation*}
			The decomposition allows us to impose shrinkage of the occurrence
			probabilities $\bm{\pi}_{kl,j}$ within the $L$ classes forming cluster
			$k$ towards a common occurrence probability vector $\bm{\mu}_{k,j}$
			for each variable $j$.  The common occurrence probability vector
			$\bm{\mu}_{k,j}$ can be interpreted as a location parameter
			characterizing the 
			distribution of the $j$th variable in the
			$k$th cluster, as the expected occurrence
			probability 
			$(\Ew{\pi_{kl,jd}})_{d=1, \ldots, D_j}$.
			
			The amount of shrinkage is determined by the precision $\phi_{k,j}$ in
			each dimension $j$, with stronger shrinkage of $\bm{\pi}_{kl,j}$
			towards $\bm{\mu}_{k,j}$ as $\phi_{k,j}$ increases.  Additionally, in
			order to avoid boundary solutions for $\bm{\alpha}_{k,j}$, a small
			regularizing constant $a_{00}>0$ is added to the decomposition:
			$ \bm{\alpha}_{k,j} = \bm{\mu}_{k,j} \phi_{k,j} +
			\alpha_{00}\bm{1}_{D_j}$, where $\bm{1}_p$ is the $p$-dimensional
			vector of ones \citep{lca:GalindoVermunt2006}.
			
			On the next level of the hierarchical prior, the prior specification
			for $\bm{\mu}_{k,j}$ needs to ensure that the component locations
			$\bm{\mu}_{k,j}$ 
			of the various clusters
			are shrunken towards an overall ``center''
			distribution in order to favor a sparse number of clusters. In the
			case of a mixture of multivariate Gaussians a straightforward option
			for a data-driven center are empirical location parameters such as the
			component-wise mean or median of the data points.  For categorical
			data, however, several possibilities might be used as ``natural''
			centers of the $\bm{\mu}_{k,j}$ distributions.  One data-driven
			possibility is to use the marginal distribution of the data. An
			alternative option is to use the uniform distribution
			$(1/D_j,\ldots,1/D_j)$ as central distribution which solely takes the
			geometry of the simplex into account.  We decided to pursue the later
			option for the following reasons: (1) given the geometry of the
			simplex it is natural to pull the cluster centers towards the centroid
			of the simplex, and (2) the likelihood contribution of an observation
			is the same regardless of the observed values for variables from the
			uniform distribution.  Thus using the uniform distribution as central
			distribution implicitly suppresses the influence of some variables in
			dominating the cluster distribution.  This eventually helps to
			estimate a sparse number of clusters.  Thus, we specify a symmetric
			Dirichlet prior on the component location $\bm{\mu}_{k,j}$ with
			hyperparameters $a_\mu$:
			\begin{align*}
				\bm{\mu}_{k,j}& \sim \mathcal{D}_{D_j}(a_\mu),
			\end{align*}
			where $\Ew{\bm{\mu}_{k,j}$ is equal to the uniform distribution and}
			$a_\mu$ controls the amount of shrinkage of the component locations
			$\bm{\mu}_{k,j}$ towards the uniform distribution. The larger the
			value $a_\mu$, the more the component occurrence probabilities are
			pulled towards the uniform distribution.
			
			Further, in order to be able to learn from the data not only the
			appropriate location but also the precision $\phi_{k,j}$ needed for each cluster
			and each dimension, we use an inverse gamma distribution as prior for
			$\phi_{k,j}$:
			\begin{align*}
				\phi_{k,j} | b_{\phi_j}& \sim \mathcal{G}^{-1}(a_\phi, b_{\phi_j}),&  b_{\phi_j}& \sim \mathcal{G}(\cb, \db), \quad j =1,\ldots,r.
			\end{align*}
			The inverse gamma prior on $\phi_{k,j}$ pulls the precision away from
			zero. This ensures a certain amount of shrinkage which is controlled
			by the parameter $b_{\phi_j}$. In order to avoid an explicit
			specification of $b_{\phi_j}$, we also impose a prior on $b_{\phi_j}$,
			i.e., a gamma distribution with fixed hyperparameters $\cb$ and $\db$.
			
\begin{figure}[t!]
				\centering
				\includegraphics[width=0.5\textwidth, clip]{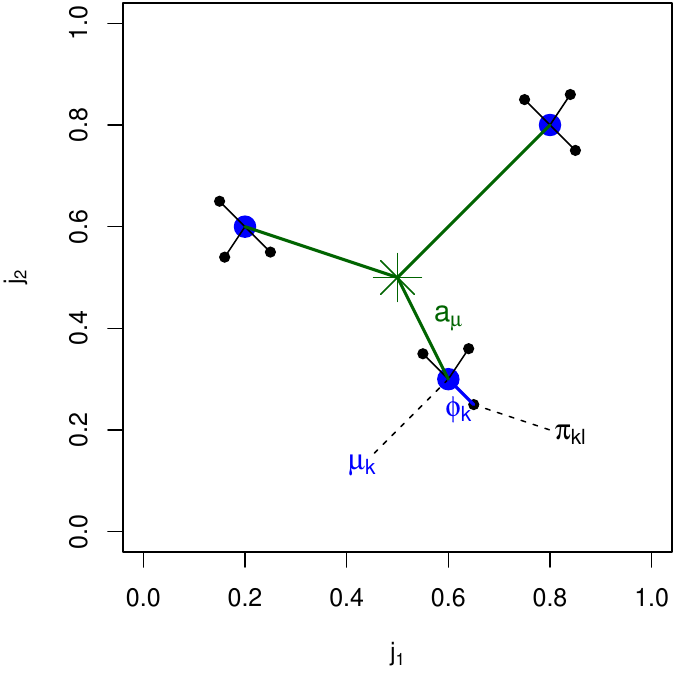}
				\caption{Schematic visualization of the hierarchical prior for a
					mixture of LCA models with three clusters, each containing three
					classes. For each cluster $k$, the occurrence probabilities
					$\bm{\pi}_{kl} =(\pi_{kl,j_1},\pi_{kl,j_2})$ of any two variables
					$j_1$ and $j_2$ in class $l$ (plotted as black points) are
					shrunken towards the cluster-centers
					$\bmu_{k}=(\mu_{k,j_1},\mu_{k,j_2})$ (plotted as large blue
					points) which in turn are shrunken towards the point
					$(1/D_j,1/D_j)=(0.5,0.5)$ (marked as a star).
				}\label{plot:MixOfMix}
			\end{figure}

			To illustrate this hierarchical prior construction, a schematic
			visualization is given in Figure~\ref{plot:MixOfMix} for two binary
			variables $j_1$ and $j_2$.  For binary variables, where $D_j=2$, the
			class occurrence probability distribution
			$\bm{\pi}_{kl,j}=(\pi_{kl,j}, 1-\pi_{kl,j})$ of variable $j$ in class
			$l$ of cluster $k$ is fully characterized by the success probability
			$\pi_{kl,j}=\Prob{Y_{j}=1}$ for category 1. Similarly, the
			cluster-specific average probability distribution
			$\bm{\mu}_{k,j} = (\mu_{k,j}, 1-\mu_{k,j})$ of the variable $j$ is
			fully characterized by $\mu_{k,j}$.
			In Figure~\ref{plot:MixOfMix}, the location occurrence probabilities
			$\bm{\mu}_{k} = (\mu_{k,j_1},\mu_{k,j_2})$ of three clusters are
			plotted as blue points. Each cluster location $\bmu_{k}$ is surrounded
			by the class occurrence probabilities
			$\bm{\pi}_{kl}= (\pi_{kl,j_1}, \pi_{kl,j_2})$ of three classes
			$l=1,2,3$ (plotted as black points).  The amount of deviation of these
			subcomponent locations from their common cluster location $\bmu_k$ is
			determined by the value of
			$\boldsymbol{\phi}_k =(\phi_{k,j_1}, \phi_{k,j_2})$.  Large values in
			$\boldsymbol{\phi}_k$ allow only for a small deviation of
			$\bm{\pi}_{kl}$ from $\bm{\mu}_{k}$ and vice versa. The hyperparameter
			$a_\mu$ controls the amount of deviation of $\bm{\mu}_{k}$ from the
			uniform distribution (here given by the point at $(0.5, 0.5)$ marked
			as a star).
			
			\subsubsection{Tuning the hyperparameters of the cluster
				distributions}\label{sec:tuning-hyperp-hier}
			
			The hierarchical prior on the occurrence probabilities
			$\bm{\pi}_{kl,j}$ needs to shrink (1) the marginal distribution of
			$\bm{\pi}_{kl,j}$ given $\bmu_{k,j}$ and 
			the hyperparameters $(a_\phi,\cb, \db)$
			towards
			the cluster-specific distribution $\bmu_{k,j}$ and (2) the
			cluster-specific distribution $\bmu_{k,j}$ towards a central
			distribution given the hyperparameter $a_{\mu}$.  This requires careful selection of the hyperparameters
			$(a_\phi,\cb, \db)$ of the inverse gamma and gamma prior as well as
			the Dirichlet hyperparameter $a_{\mu}$
			in the prior for  $\bmu_{k,j}$
			which are instrumental for
			learning the subcomponent occurrence probabilities $\bm{\pi}_{KL}$
			jointly with the parameters
			$\bm{\mu}_K=(\bm{\mu}_{k,j})_{j=1, \ldots, r;k=1, \ldots,K}$ and
			$\bm{\phi}_K=(\phi_{k,j})_{j=1, \ldots, r;k=1, \ldots,K}$ from the
			data.
			
\begin{figure}[t!]
				\centering
				\includegraphics[width=\textwidth, trim = 0 10 0 0]{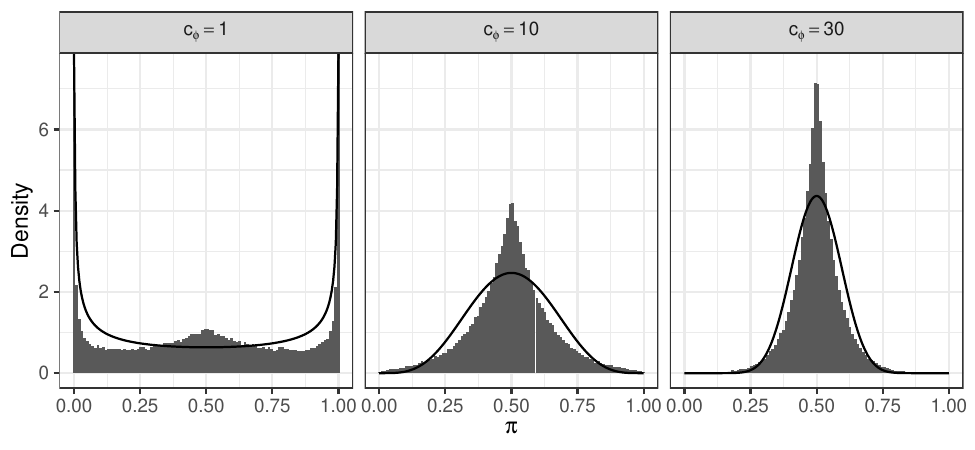}
				\caption{Marginal prior distribution
					$p(\pi_{kl,j}| \mu_{k,j}, a_\phi,\cb,\db) $ of the success
					probability $\pi_{kl,j}$ of category 1 for a binary variable $j$
					with $D_j = 2$, assuming a beta distribution
					$\pi_{kl,j} \sim \mathcal{B}(\mu_{k,j} \phi_j,
					(1-\mu_{k,j}) \phi_j)$ with $\mu_{k,j} =0.5$ for the
					success probability of category 1 in the cluster-specific average
					probability distribution
					$\bm{\mu}_{k,j} = (\mu_{k,j}, 1-\mu_{k,j})$,
					$\phi_j\sim \cG^{-1}(a_\phi,b_{\phi_j})$, and
					$ b_{\phi_j} \sim \cG(\cb,\db)$ for $a_\phi=\cb=\db=1$ (left),
					$a_\phi=1, \cb=10, \db=1$ (middle) and $a_\phi=1, \cb=30, \db=1$
					(right).  For comparison, the black line shows the beta
					distribution for $\pi_{kl,j}$ with the same mean and
					variance as the marginal prior distribution.}
				\label{plot:invgam}
			\end{figure}

			Specifying a supposedly uninformative prior for $\bm{\pi}_{kl,j}$ with
			$a_\phi=\cb=\db=1$ leads to a diffuse marginal prior distribution
			$p(\bm{\pi}_{kl,j}| \bm{\mu}_{k,j}, a_\phi,\cb,\db) $ which supports
			not only probabilities around the cluster-specific distribution
			$\bm{\mu}_{k,j}$, but also extreme probabilities close to 0 and 1.
			This can be seen in Figure~\ref{plot:invgam} on the left-hand side,
			where the marginal prior distribution
			$p(\pi_{kl,j}| \mu_{k,j}, a_\phi,\cb,\db) $ of the success probability
			$\pi_{kl,j}$ of category 1 for binary data with $D_j = 2$ is plotted
			for $a_\mu=0.5$ (obtained using simulation).
			However, these extreme probability values close to 0 and
			1 need to be avoided as they do not support the modeling aim of
			pulling the classes together by shrinking the occurrence probabilities
			towards the common component mean.  In contrast, setting $\cb\gg 1$
			leads to the desired shrinkage shape of the marginal occurrence
			probabilities, see Figure~\ref{plot:invgam} in the middle and on the
			right-hand side.

			Note that specifying a standard Dirichlet prior
			$\bm{\pi}_{kl,j} \sim \mathcal{D}_{D_j}(\bm{\alpha})$ with a fixed
			hyperparameter vector $\bm{\alpha}_{k,j}=\bm{\alpha}$ on the
			occurrence probabilities $\bm{\pi}_{kl,j}$ would also favor
			probabilities around a pre-specified mean value, if the sum over the
			elements in $\bm{\alpha}$ is large.  However, compared to the marginal
			distribution obtained by decomposing the hyperparameter
			$\bm{\alpha}_{k,j}$ and specifying a hierarchical prior, less
			shrinkage towards the mean would be induced because of the platicurtic
			shape of the Dirichlet distribution, see Figure~\ref{plot:invgam} in
			the middle and on the right-hand side, where the density of a beta
			distribution having the same mean and variance as the proposed
			marginal distribution is plotted with a solid black line.
			
			In the following we fix $a_{\phi} = \db = 1$ and only investigate in
			detail how different choices for $a_{\mu}$ and $\cb$ impact the
			results obtained. These hyperparameter values have a crucial influence
			on the amount of shrinkage and provide sufficient flexibility, even if
			$a_{\phi}$ and $\db$ are fixed at 1.
			
			\section{Posterior inference}\label{sec:posterior-inference}
			
			\subsection{MCMC estimation}\label{sec:mcmc-estimation}
			
			Bayesian estimation of the mixture of LCA models is performed using
			Markov chain Monte Carlo (MCMC) sampling with data augmentation.  
			First, we consider the number of components $K$ a latent variable which is sampled during MCMC. Given $K$, the
			component and subcomponent assignments on the upper and lower level,
			$\bS=(S_1,\ldots,S_N)$ and $\bI = (I_1,\ldots,I_N)$, respectively, are
			added as latent variables and also drawn during MCMC sampling.
			Specifically, $S_i \in \{1,\ldots,K\}$ assigns each observation
			$\by_i$ to cluster $S_i$ on the upper level of the mixture of LCA
			models. On the lower level, $I_i \in \{1,\ldots,L\}$ assigns
			observation $\by_i$ to subcomponent $I_i$. Hence, the pair
			$(S_i, I_i)$ carries all the information needed to assign each
			observation to a unique class in the two-layer mixture.
			
			In each iteration of the MCMC sampling scheme, the latent component
			assignments $(S_1,\ldots,S_N)$ on the upper level induce a random
			partition of the data, i.e., two observations $\by_i$ and $\by_j$
			belong to the same cluster if and only if $S_i=S_j$. Thus the sampling
			scheme directly provides the posterior distribution of the partitions
			$\cC = \{C_1,\ldots,C_{K_+}\}$, with $C_k=\{i:S_i=k\}$ being the index
			set of observations assigned to cluster $k$, and $K_+$ the induced
			number of data clusters.
			
			In order to obtain samples of $K$, and, conditional on $K$, of
			$(\bS, \bm{\eta}_K, \bm{w}_{KL},
			\bm{\mu}_K,\bm{\phi}_K,\bI,
			\bm{\pi}_{KL},\alpha,\bm{b}_\phi)$ from the posterior distribution,
			given data $\bm{y}=(\bm{y}_1, \ldots, \bm{y}_N)$, a transdimensional
			sampler is required which is able to sample parameter vectors of
			varying dimension. We use the telescoping sampler proposed by
			\cite{Fruehwirth-Schnatter+Gruen+Malsiner-Walli:2021}.  This MCMC
			sampling scheme includes a sampling step where $K$ is explicitly
			sampled as an unknown parameter, but otherwise requires only sampling
			steps of a finite mixture model with a fixed number of components.
			
			In particular, the telescoping sampler 
				distinguishes explicitly 
			between $K$, the number of components in the mixture distribution, and
			$K_+$, the number of ``filled'' components, i.e., components to which
			observations are assigned and which actually correspond to data
			clusters.  Updating of the number of data clusters $K_+$ is implicitly
			performed, based on the sampled partition. In contrast, the number of
			components $K$ is explicitly sampled from the conditional posterior of
			$K$, given the current partition $\cC$ which is characterized by the
			number of clusters $K_+$ and the number of observations
			$N_k=\#\{i|S_i=k\}$ assigned to each cluster $k=1, \ldots, K_+$.  
				Using a dynamic specification for the Dirichlet prior on the
				weights, 
			this posterior is proportional to the conditional
			partition distribution times the prior on $K$, i.e.,
			\begin{align}\label{eq:postK}
				p(K|\cC,\alpha) &\propto p(\cC|\alpha,K)  p(K) \\
				&\propto \frac{K!}{(K-K_+)!}
				\frac{\Gamma(\alpha)}{\Gamma(\alpha +N)}	
				\prod_{k=1}^{K_+} \frac{\Gamma(N_k+ \alpha/K)}{\Gamma(1+ \alpha/K)} p(K), \nonumber
			\end{align}
			where $\alpha$ is the hyperparameter of the symmetric Dirichlet prior
			$\bm{\eta}_K|K,\alpha \sim \mathcal{D}_K(\alpha/K)$.  Importantly,
			sampling $K$ only depends on the current partition $\cC$, and is
			independent of the component parameters.  This makes the sampling
			scheme a generic sampler for arbitrary component densities and
			particularly useful for mixtures with complex component
			densities. More details on the telescoping sampler can be found in
			\cite{Fruehwirth-Schnatter+Gruen+Malsiner-Walli:2021}.
			
			Conditional on $K$, sampling of the other parameters is
			straightforward as one can directly build on the standard sampling
			scheme for a mixture with fixed $K$.  Gibbs updates based on the
			Dirichlet distribution are possible for the component and subcomponent
			weights as well as for the subcomponent occurrence probabilities on
			the lower level.  For the parameters on the cluster level,
			Metropolis-Hastings steps are required to draw $\bm{\mu}_{k,j}$ and
			$\phi_{k,j}$ using a Dirichlet proposal for $\bm{\mu}_{k,j}$ and a
			random walk proposal for the logarithm of $\phi_{k,j}$. Updates on
			$\alpha$ are also generated by a Metropolis-Hastings step with a
			log-normal random walk proposal.  Parameters for newly created,
			empty components are sampled from the priors. Sampling of
			hyperparameters is only based on information from the filled
			components. The MCMC sampling scheme including the initialization
			strategy pursued is reported in detail in
			Appendix~\ref{sec:mcmc-scheme}.
			
			Due to multimodality of the posterior and mixing issues, inference is
			in the following based on 10 parallel runs.  For each run, 4,000
			iterations are recorded after discarding the first 1,000 iterations as
			burn-in.  A specific run for further posterior inference is determined
			in the following way: First, the number of clusters $K_+$ is estimated
			by taking the most frequent mode of the posterior $p(K_+|\bm{y})$
			across the 10 parallel runs.  Then, among the runs where the mode
			corresponds to the selected $K_+$, the run with the highest mixture
			likelihood, is selected and used for further posterior inference.
			
			\subsection{Resolving label switching to obtain a final clustering}\label{sec:resolv-label-switch}	
			
			The draws from the posterior distribution require post-processing to
			resolve the label switching issue on the upper level to estimate a
			final clustering of the data. 
				Many useful techniques have been
				proposed to post-process the sampled partitions \citep[see,
				e.g.,][]{lca:Papastamoulis:2016,lca:Wade+Ghahramani:2018,lca:Dahl+Johnson+Mueller:2022}. However,
				we would like to not only obtain a posterior estimate of a partition
				of the data but also perform posterior inference for the
				cluster-specific distributions, i.e., obtain the posterior
				distribution of the component-specific parameters.  We adapt the
				procedure considered in
				\citet{Malsiner-Walli+Fruehwirth-Schnatter+Gruen:2017} to identify
				the model on the upper level.  The procedure consists of the
				following steps. 
			
			First, a model selection
			step is performed by estimating the number of data clusters using the
			mode $\hat{K}_+$ of the posterior $p(K_+|\bm{y})$, i.e., by selecting
			the most frequent number of filled components during MCMC sampling.
			Only draws corresponding to iterations with exactly $\hat{K}_+$ filled
			components are included in the further analysis.  Then, for each
			iteration and filled component, a low-dimensional functional of the
			class-specific parameters is computed to characterize the components.
			We use the averaged class occurrence probabilities weighted by their
			class weights as functional, i.e.,
			$\tilde{\bm{\pi}}_{k,j} =\sum_{l=1}^L w_{kl}\bm{\pi}_{kl,j}$.  These
			functionals are stacked on top of each other for the different
			components and the rows in the resulting matrix are clustered into
			$\hat{K}_+$ clusters using $k$-means clustering.
			
\begin{figure}[t!]
				\centering
				\includegraphics[width=\textwidth, trim = 0 12 0 5, clip]{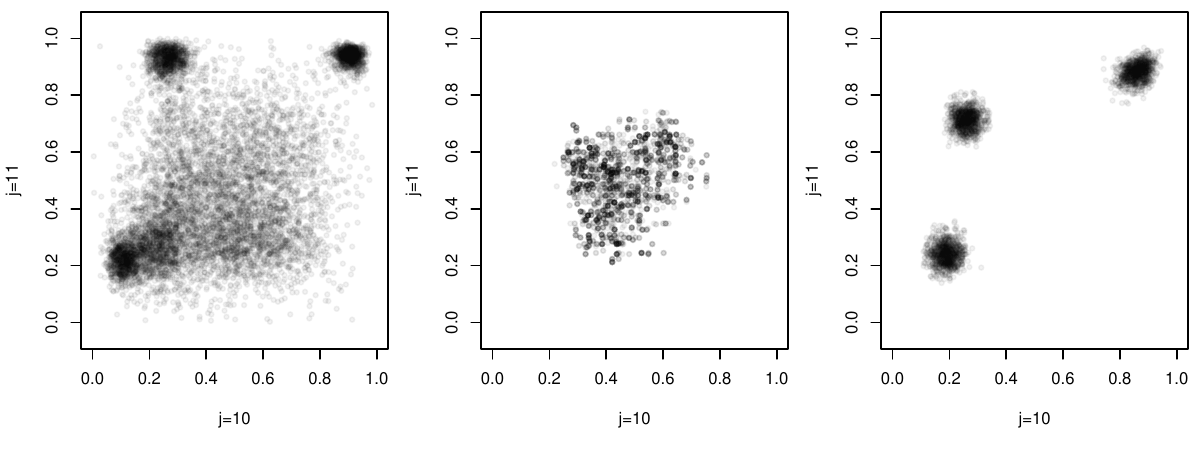}
				\caption{Simulation study, $\rho=0.3$. Diagnostic plot
					of one MCMC
					run of a simulated data set with $L=3$, $a_\mu=20$, $\cb=30$,
					1,000 iterations (after a burn-in of 1,000 iterations).  For two
					variables ($j \in \{10, 11\}$), scatter plots of the sampled
					$\bm{\pi}_{kl,j}$ (left),
					$\bm{\mu}_{k,j}$
					(middle) and the
					cluster-specific functional
					$\tilde{\bm{\pi}}_{k,j}=\sum_{l=1}^L w_{kl}
					\bm{\pi}_{kl,j}$ (right) are shown. }\label{plot:ppr}
			\end{figure}

			Using a functional of the class-specific parameters eases capturing
			the clusters in the parameter space. This is illustrated in
			Figure~\ref{plot:ppr}. For one run of the simulation study in
			Section~\ref{sec:empir-demonstr} with three classes ($L=3$) on the lower
			and three
			filled components ($K_+=3$) 
				on the upper level, Figure~\ref{plot:ppr} shows scatter plots
			of the sampled class occurrence probabilities $\bm{\pi}_{kl,j}$ for
			all three classes within cluster $k$ (left-hand side), the cluster
			locations $\bmu_k$ (middle) and the functional
			$\tilde{\bm{\pi}}_{k,j}$ (right-hand side) for all three clusters
			$k=1,2,3$.
			Figure~\ref{plot:ppr} on the left indicates that the class occurrence
			probabilities in the various clusters differ in their variability. In
			particular the rather diffuse subcomponents show an overlap within as
			well as across clusters, i.e., model identification is difficult if
			based directly on the class occurrence probabilities
			$\bm{\pi}_{kl,j}$.  By contrast, the sampled component locations
			$\bm{\mu}_{k,j}$ are shrunken towards the center and exhibit
			considerable overlap, making identification based on $\bm{\mu}_{k,j}$
			difficult (Figure~\ref{plot:ppr} in the middle). However, a clear
			separation can be seen for the values of the functional
			$\tilde{\bm{\pi}}_{k,j}$ induced by the averaged class occurrence
			probabilities (Figure~\ref{plot:ppr} on the right). These functionals
			thus represent the most suitable parameters for identifying the three
			clusters which clearly is an easy task based on these quantities.
			
			Now $k$-means clustering is employed to assign each functional of the
			same iteration to a different group, inducing a unique labeling of the
			functional values. This unique labeling is used to reorder the
			components for all considered iterations thus obtaining identified
			values for the cluster weights $\boldeta_k$, the cluster locations
			$\bmu_{k,j}$ and the cluster allocations $S_i$.  A final partition of
			the data is obtained based on the maximum a posteriori (MAP) estimates
			of the relabeled component allocations $S_i$.
			
			Note that on the lower level, in contrast to the upper level,
			identification of the classes is not required, as the LCA model is
			only intended to provide a semi-parametric approximation of the
			cluster distribution by capturing a possible association structure
			within the cluster. Thus, label switching among classes of the same
			component does not impact on the identification of the cluster
			distribution on the upper level and can be ignored, since the
			functional $\tilde{\bm{\pi}}_{k,j}$ is invariant to label switching
			among the $L$ classes in any cluster $k$.
			
			\section{Empirical demonstrations}\label{sec:empir-demonstr}
			\subsection{Simulation study}\label{sec:sim}
			
			The main focus of the simulation study is (1) to investigate whether
			the proposed Bayesian mixture of LCA models approach is able to
			estimate the true number of data clusters and the true partition of
			the data when the variables are associated within the clusters and (2)
			to contrast the performance to standard approaches for clustering
			multivariate categorical data based on the conditional independence
			assumption within clusters.  We also investigate the influence of the
			specified number of subcomponents $L$ on the clustering results and
			perform a sensitivity analysis of the impact of the specified amount
			of shrinkage imposed by the chosen hyperparameters of the hierarchical
			prior on the occurrence probabilities.
			
			\begin{table}[t!]
				\centering
				\caption{Simulation study. Occurrence probabilities
					$\Prob{Y_{ij}=1|S_i=k}$ for clusters $k=1,2,3$ and variables
					$j=1,\ldots,30$.}\label{tab:prob}
				\begin{tabular}{rrrrrrrrrr}
					\toprule
					& V1 & $\ldots$ & V10 & V11 & $\ldots$ & V20 & V21 & $\ldots$ & V30 \\
					\midrule
					cluster 1 & 0.8 & $\ldots$ & 0.8 & 0.8 & $\ldots$ & 0.8 & 0.2 & $\ldots$ & 0.2 \\
					cluster 2 & 0.2 & $\ldots$ & 0.2 & 0.8 & $\ldots$ & 0.8 & 0.2 & $\ldots$ & 0.2 \\
					cluster 3 & 0.2 & $\ldots$ & 0.2 & 0.2 & $\ldots$ & 0.2 & 0.8 & $\ldots$ & 0.8  \\
					\bottomrule
				\end{tabular}
			\end{table}
			
\begin{figure}[t!]
				\includegraphics[width=\textwidth]{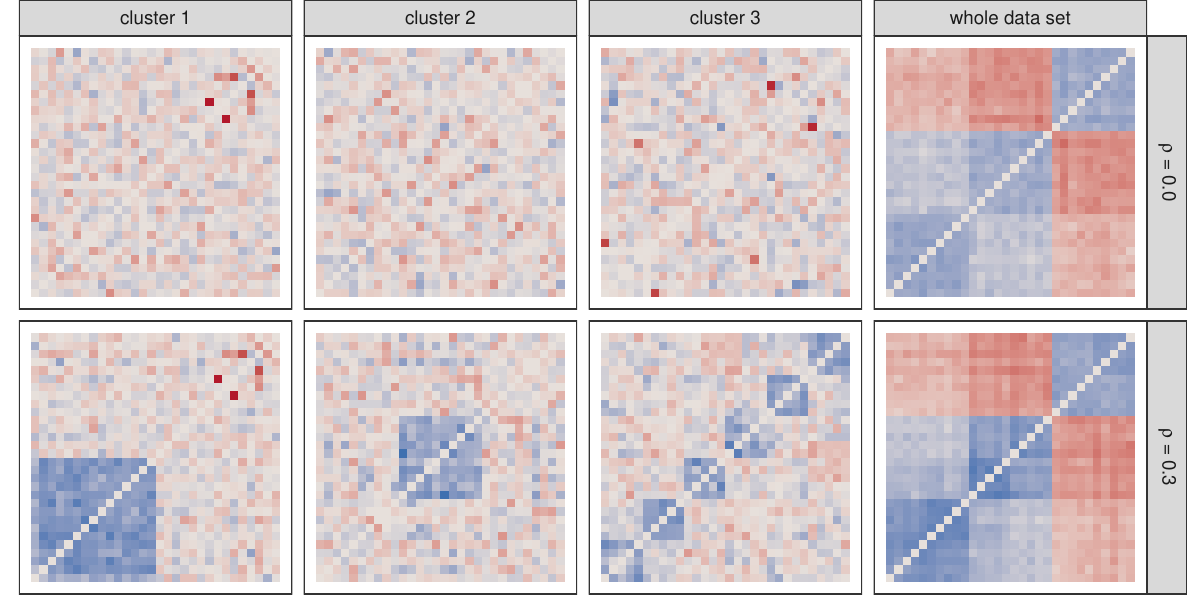}
				\caption{Simulation study. Top row:
					$\rho = 0.0$;
					bottom row:
					$\rho = 0.3$.
					Empirical results for one data set with $N = 500$ of
					the log-odds ratios of the variables, conditional on cluster 1, 2
					and 3 respectively and marginally across the whole data set (from
					left to the right).}\label{plot:simdata1}
			\end{figure}

			Inspired by the low back pain data set (see
			Section~\ref{sec:low-back-pain}), we define a mixture with three
			clusters of multivariate distributions on 30 binary variables as data
			generating mechanism. We consider two different scenarios regarding
			the within-cluster association between variables, one with associated
			variables and one where variables are independent.  For both
			scenarios, the same marginal cluster-specific occurrence probabilities
			are assumed, as given in Table~\ref{tab:prob}.  In the first scenario,
			the within-cluster associations are generated by sampling from a
			multivariate discrete distribution with a given correlation
			matrix. The defined correlation matrix contains blocks of variables
			where within blocks all variables are pairwise correlated with
			correlation $\rho=0.3$.  Outside the blocks the variables are
			independent.  The three clusters differ in regard to location and size
			of the blocks in the correlation matrix. In cluster~1 and 2 only one
			block is present which consists of the first 15 and the middle 10
			variables, respectively. Cluster~3 contains six blocks of successive
			variables. The association structure of this setup is visualized in
			Figure~\ref{plot:simdata1} in the bottom row, where for one data set
			the matrices containing the log-odds ratios of the success
			probabilities of the variables for each cluster and also for the whole
			data set are shown. The blocks of associated variables can be clearly
			identified through very high/low log-odds ratios.
			
			The \textsf{R} \citep{lca:RCore:2024} package \pkg{GenOrd}
			\citep{lca:Barbiero+Ferrari:2015} is used to generate 30 data sets
			with 500 observations. 
				The three clusters are specified to have
				the same group size. 
			Package \pkg{GenOrd} employs a copula-based
			procedure for generating samples from discrete random variables with a
			prescribed correlation matrix and marginal distributions. Notably the
			association structure in the clusters are not generated by a LCA
			model. This means that the data generating process of the simulated
			data deviates from the modeling approach. 
				Using this setting we
				show that the proposed approach is able to capture the association
				within a cluster, even if the cluster is not generated by a LCA
				model. 
			
			To each data set, a Bayesian mixture of LCA models is fitted with
			prior distributions and hyperparameters as described in
			Section~\ref{sec:prior-specification}. In particular, we use
			$K\sim \mathit{BNB}(1,4,3)$, $\alpha\sim \cG(1,2)$, $\delta=1$,
			$a_\phi=\db=1$.  To investigate the impact of the hyperparameters
			$a_\mu$ and $\cb$ and the number of classes $L$ forming one cluster, a
			sensitivity analysis is performed. Specifically, the combination of
			the following values is investigated: $a_\mu\in(1,10,20,30)$,
			$\cb\in(1,10,20,30,40)$ and $L\in(2,3,4,5)$.
			
			For each combination of prior parameter values, the specified model is
			fitted to each of the 30 data sets. As described in
			Section~\ref{sec:mcmc-estimation}, for each data set, 10 parallel runs
			with 4,000 recorded iterations after discarding the first 1,000
			iterations as burn-in are performed and, using the $K_+$ value
			estimated by the most frequent mode, the run with the highest mixture
			likelihood is used for further posterior inference. 
				Convergence of the chains was assessed by visual inspection of the trace plots. 
			Clustering
			performance is measured using the adjusted Rand index (ARI) which
			measures the classification agreement between the true and the
			estimated cluster memberships corrected for agreement by chance given
			the marginal cluster sizes. An ARI close to zero will be obtained for
			two independent partitions while an ARI close to 1 indicates perfect
			agreement between the two partitions.
			
			Finally, we compare our results to the clustering results obtained by
			alternative methods. In particular, we use the package \pkg{LCAvarsel}
			\citep{lca:Fop:2018} which combines LCA modeling with variable
			selection to omit redundant variables inducing association with the
			cluster distributions. In addition, we also perform standard latent
			class analysis based on maximum likelihood estimation with an
			expectation-maximization algorithm using the \textsf{R} packages
			\pkg{BayesLCA} \citep{lca:White+Murphy:2014}, \pkg{Rmixmod}
			\citep{lca:Lebret+Iovleff+Langrognet:2015} and \pkg{poLCA}
			\citep{lca:Linzer+Lewis:2011}.  Model selection is performed based on
			the Bayesian information criterion (BIC; \pkg{BayesLCA}, \pkg{poLCA})
			and ICL (\pkg{Rmixmod}), after having fitted models with an increasing
			number of $K$ from 2 to 10 to all data sets and the packages are
			otherwise used with their default options.  As an alternative Bayesian
			approach, we use package \pkg{PReMiuM} \citep{FM:Liverani2015} to fit
			an LCA model with a Dirichlet process prior.  The number of initial
			clusters is set to 30, the number of burn-in draws and number of
			iterations are set to 2,000 and 10,000, respectively. All other
			settings, such as hyperparameter specifications, calculation of the
			similarity matrix and the derivation of the best partition, are left
			at the default values of the package.  Additionally, \pkg{PReMiuM}
			with variable selection is performed using either the approach by
			\cite{lca:Papathomas2012} (PReMiuM-contVS) or the method proposed by
			\cite{lca:ChungDunson2009} (PReMiuM-binVS).
			
			Table~\ref{tab:simdata1} reports the clustering results obtained using
			the proposed Bayesian mixture of LCA models approach for each
			combination of prior hyperparameter values investigated.  The average
			results obtained over the 30 data sets are reported.  As expected,
			imposing a considerable amount of shrinkage supports the estimation of
			the true number of clusters $K^{\text{true}}=3$. In particular,
			specifying $a_\mu\ge 10$ and $\cb \ge 20$ leads to the estimation of
			three data clusters for almost all data sets. The specification of a
			larger number of classes $L$ also helps to obtain a sparser cluster
			solution. Results for $L = 2$ indicate that two classes might not be
			sufficient to capture the association structure within the three
			clusters of the true data generating process and thus more than three
			clusters are frequently estimated, even if the prior parameter values
			induce a considerable amount of shrinkage. In case $L \ge 3$, results
			are comparable regardless of the specific value. This implies that
			$L = 3$ seems to be sufficient to capture the association within the
			clusters in a suitable way and adding more classes does not improve
			but also does not lead to a deterioration of results. For
			hyperparameter values inducing a suitable amount of shrinkage as well
			as for $L \ge 3$, the average ARI is between 0.72 and 0.78. This value
			is clearly higher than the ARI of the clustering solutions obtained
			with the standard methods, see Table~\ref{tab:others} on the
			left. Note that for the standard methods most of the time too many
			clusters are estimated (between four and eight), with the lowest
			number estimated for the variable selection approach proposed by
			\citet{lca:FopetAl2017}.
			
			Overall, the clustering results of the Bayesian mixture of LCA models
			approach indicate that -- at least for the investigated simulation
			scenario -- the clustering results are robust in regard to moderate
			changes in the specification of the shrinkage parameters and the
			number of classes $L$, as long as they are chosen sufficiently large.
			
\begin{figure}[t!]
						\centering
						\includegraphics[width=\textwidth]{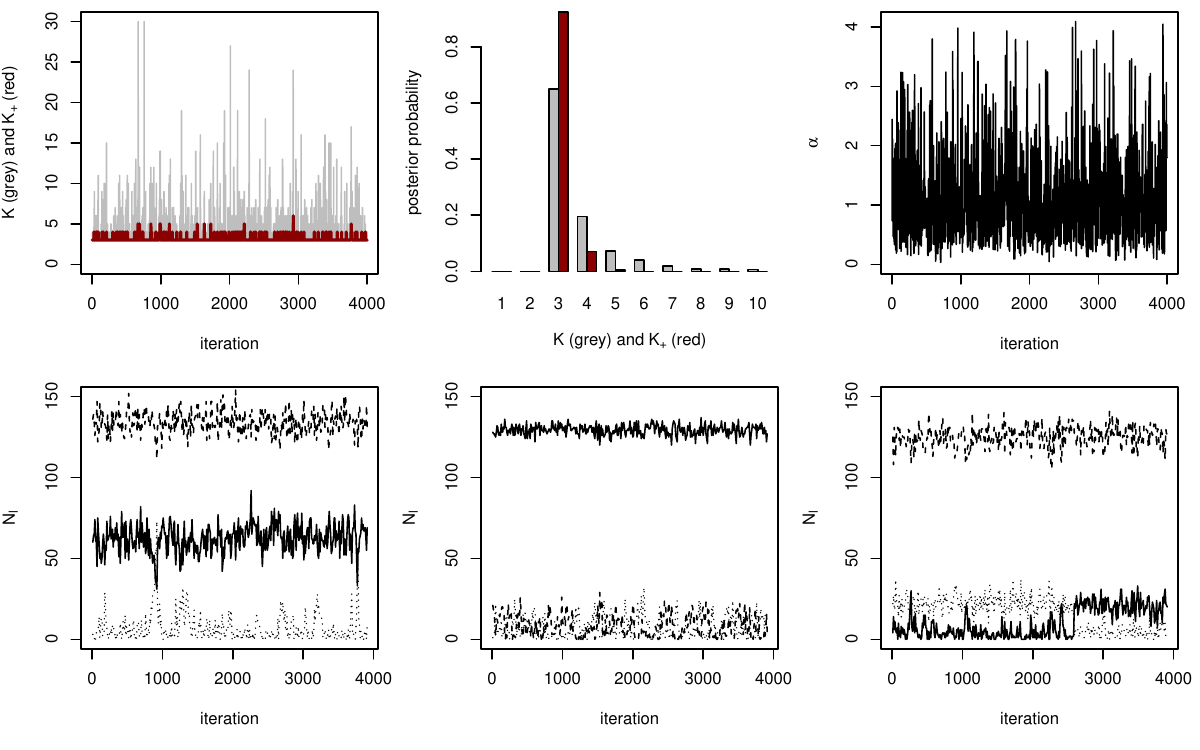}
						\caption{Simulation study, $\rho=0.3$. Diagnostic plot of one MCMC
							run of a simulated data set with $L=3$, $a_\mu=20$, $\cb=30$,
							4,000 iterations (after a burn-in of 1,000 iterations). Top: trace
							plot (left) and posterior distribution (middle) of the number of components $K$ (gray) and number of data
							clusters $K_+$ (red), 
								sampled hyperparameter $\alpha$ (right). 
							Bottom: trace plot of the number of
							observations assigned to subcomponent $l=1,2,3$ for cluster 1, 2
							and 3 (from left to right).}\label{plot:diag}
					\end{figure}

			Figure~\ref{plot:diag} provides detailed insights into the specific
			MCMC results obtained for one run using $L=3$. On the top left, the
			trace plot of the sampled $K$ and the induced number of clusters $K_+$
			is shown.  Values considerably larger than three are sampled for $K$;
			the number of filled components $K_+$ switches between three and five,
			but are equal to $K_+=3$ most of the time. The barplots in
			Figure~\ref{plot:diag} 
				in the middle indicate that both posterior
				distributions $p(K|\by)$ and $p(K_+|\by)$ have their mode at
				three. 
					The trace plot on the top right hand side indicates
					convergence of the chain for the sampled $\alpha$ values.
					The
					trace plots at the bottom of Figure~\ref{plot:diag} show the number of
					observations allocated to the classes belonging to cluster~1, 2, and
					3, respectively. In cluster~1 and 3 (first and third plot), the sizes
					of the three classes are quite distinct. In cluster~2 (second plot),
					there are two small classes where the number of observations assigned
					varies a lot. This suggests that perhaps at least one class is
					redundant. However, overfitting $L$ is allowed and does not have any
					impact on the cluster distribution. Scatter plots of the sampled class
					probabilities $\pi_{kl,j}$ have been shown in Figure~\ref{plot:ppr}.
					
					\newcommand{\figcapADD}[1]{. The true number of clusters is equal to 3,
						the number of simulated data sets is equal to 30.}
					
					\begin{table}[t!]
						\centering
						\caption{Simulation study. Clustering results (number of clusters
							$K_+$ and ARI) for various specifications of $a_\mu$, $\cb$, and
							$L$. $N=500$, $r=30$, correlation $\rho=0.3$\figcapADD . }\label{tab:simdata1}
						\begin{tabular*}{\textwidth}{@{\extracolsep\fill}rr|rrrrr|rrrrr}
							\toprule
							\multicolumn{2}{c}{}& \multicolumn{5}{c}{$K_+$}&\multicolumn{5}{c}{ARI}\\
							\cmidrule{3-7}\cmidrule{8-12}%
							&&\multicolumn{5}{c}{$\cb$} &\multicolumn{5}{c}{$\cb$}\\
							& $a_\mu$ & 1 & 10 & 20 & 30 & 40 & 1 & 10 & 20 & 30 & 40 \\
							\midrule
							$L=2$ & 2 & 7.00 & 5.80 & 6.20 & 6.10 & 6.20 & 0.57 & 0.64 & 0.64 & 0.65 & 0.64 \\
							& 10 & 7.00 & 4.50 & 3.60 & 3.40 & 3.40 & 0.54 & 0.67 & 0.75 & 0.76 & 0.77 \\
							& 20 & 6.90 & 4.20 & 3.50 & 3.30 & 3.10 & 0.52 & 0.65 & 0.71 & 0.75 & 0.76 \\
							& 30 & 7.20 & 4.20 & 3.60 & 3.40 & 3.30 & 0.51 & 0.62 & 0.69 & 0.73 & 0.74 \\	
							\midrule
							$L=3$ & 2 & 5.90 & 5.20 & 5.40 & 5.60 & 5.90 & 0.62 & 0.65 & 0.67 & 0.65 & 0.64 \\
							& 10 & 5.40 & 3.70 & 3.20 & 3.10 & 3.10 & 0.57 & 0.72 & 0.77 & 0.78 & 0.77 \\
							& 20 & 5.60 & 3.80 & 3.30 & 3.10 & 3.10 & 0.54 & 0.67 & 0.73 & 0.75 & 0.76 \\
							& 30 & 5.50 & 3.70 & 3.30 & 3.20 & 3.10 & 0.51 & 0.63 & 0.71 & 0.74 & 0.75 \\
							\midrule
							$L=4$ & 2 & 5.10 & 4.60 & 5.00 & 5.30 & 5.80 & 0.65 & 0.67 & 0.66 & 0.64 & 0.65 \\
							& 10 & 4.60 & 3.30 & 3.10 & 3.00 & 3.10 & 0.62 & 0.75 & 0.78 & 0.78 & 0.77 \\
							& 20 & 4.80 & 3.50 & 3.20 & 3.10 & 3.00 & 0.55 & 0.68 & 0.75 & 0.77 & 0.77 \\
							& 30 & 5.00 & 3.40 & 3.30 & 3.20 & 3.10 & 0.53 & 0.66 & 0.72 & 0.74 & 0.75 \\
							\midrule
							$L=5$ & 2 & 4.70 & 4.60 & 4.90 & 5.20 & 5.40 & 0.68 & 0.71 & 0.70 & 0.66 & 0.64 \\
							& 10 & 4.20 & 3.10 & 3.00 & 3.10 & 2.90 & 0.62 & 0.76 & 0.78 & 0.77 & 0.75 \\
							& 20 & 4.20 & 3.30 & 3.10 & 3.00 & 3.00 & 0.55 & 0.72 & 0.76 & 0.77 & 0.77 \\
							& 30 & 4.40 & 3.40 & 3.20 & 3.20 & 3.10 & 0.53 & 0.68 & 0.73 & 0.74 & 0.75 \\
							\bottomrule
						\end{tabular*}
					\end{table}
					
					\begin{table}[t!]
						\caption{Simulation study. Clustering results (number of clusters
							$K_+$, ARI and err) for alternative methods.  $N= 500$, $r=30$,
							correlation $\rho=0.3$ (left) and $\rho = 0.0$ (right)\figcapADD .}\label{tab:others}
						\begin{tabular*}{\textwidth}{@{\extracolsep\fill}r rrr rrr}
							\toprule
							& \multicolumn{3}{@{}c@{}}{$\rho=0.3$} & \multicolumn{3}{@{}c@{}}{$\rho=0.0$} \\
							\cmidrule{2-4}\cmidrule{5-7}%
							& $K_+$ & ARI & err & $K_+$ & ARI & err \\
							\midrule
							LCAvarsel & 3.43 & 0.67 & 0.15 & 3.00 & 0.96 & 0.02\\
							BayesLCA & 4.07 & 0.68 & 0.21 & 3.00 & 0.96 & 0.02 \\
							PoLCA & 4.17 & 0.67 & 0.21 & 3.03 & 0.95 & 0.02 \\
							Rmixmod & 7.07 & 0.68 & 0.19 & 3.20 & 0.95 & 0.02 \\
							PReMiuM & 8.67 & 0.52 & 0.36 & 3.13 & 0.96 & 0.02 \\
							PReMiuM-contVS & 7.93 & 0.56 & 0.31 & 3.07 & 0.95 & 0.02 \\
							PReMiuM-binVS & 8.23 & 0.54 & 0.34 & 3.17 & 0.95 & 0.02 \\
							\bottomrule
						\end{tabular*}
					\end{table}
					
					The aim of the second simulation scenario is to investigate how the
					specified Bayesian mixture of LCA models performs if variables within
					clusters are independent, i.e., if the standard conditional
					independence assumption of the LCA model applies to the cluster
					distributions. The data generating process is basically the same as
					for the first scenario, the only difference is that the pairwise
					correlation in blocks is set to $\rho=0.0$. The matrices showing the
					pairwise log-odds within clusters and for the whole data set are
					provided in Figure~\ref{plot:simdata1} in the top row.
					
					Table~\ref{tab:simdata2} reports the clustering results of the
					proposed Bayesian mixture of LCA models approach for each combination
					of prior parameter values investigated. In the case of independent
					clusters, the Bayesian modeling approach consistently estimates three
					clusters with an ARI around 0.95 for all prior specifications which
					also performed well in the simulations scenario with correlated
					clusters. These results also coincide with the favorable results
					obtained with the standard methods, as shown in Table~\ref{tab:others}
					on the right. The good performance of the Bayesian mixture of LCA
					models approach also in the case where its flexibility is indeed not
					required implies that it can be employed even if the user is unsure
					about the dependency structure within the clusters they are aiming
					for.
					
					\begin{table}[t!]
						\centering
						\caption{ Simulation study. Clustering results (number of clusters
							$K_+$ and ARI) for various specifications of $a_\mu$, $\cb$, and
							$L$. $N=500$, $r=30$, correlation $\rho=0.0$\figcapADD .}\label{tab:simdata2}
						\begin{tabular*}{\textwidth}{@{\extracolsep\fill}rr|rrrrr|rrrrr}
							\toprule
							\multicolumn{2}{c}{}& \multicolumn{5}{c}{$K_+$}&\multicolumn{5}{c}{ARI}\\
							\cmidrule{3-7}\cmidrule{8-12}%
							&&\multicolumn{5}{c}{$\cb$} &\multicolumn{5}{c}{$\cb$}\\
							& $a_\mu$ & 1 & 10 & 20 & 30 & 40 & 1 & 10 & 20 & 30 & 40 \\
							\midrule
							$L=2$ & 2 & 3.00 & 3.00 & 3.00 & 2.90 & 2.80 & 0.96 & 0.95 & 0.95 & 0.92 & 0.87 \\
							& 10 & 3.00 & 3.00 & 3.00 & 3.00 & 3.00 & 0.95 & 0.96 & 0.96 & 0.96 & 0.95 \\
							& 20 & 3.00 & 3.00 & 3.00 & 3.00 & 3.00 & 0.95 & 0.95 & 0.95 & 0.96 & 0.95 \\
							& 30 & 3.00 & 3.00 & 3.00 & 3.00 & 3.00 & 0.95 & 0.95 & 0.95 & 0.95 & 0.95 \\
							\midrule
							$L=3$ & 2 & 3.00 & 3.00 & 2.90 & 2.70 & 2.60 & 0.95 & 0.94 & 0.89 & 0.84 & 0.80 \\
							& 10 & 3.00 & 3.00 & 3.00 & 3.00 & 3.00 & 0.95 & 0.95 & 0.96 & 0.95 & 0.95 \\
							& 20 & 3.00 & 3.00 & 3.00 & 3.00 & 3.00 & 0.95 & 0.95 & 0.96 & 0.96 & 0.96 \\
							& 30 & 3.00 & 3.00 & 3.00 & 3.00 & 3.00 & 0.96 & 0.96 & 0.95 & 0.95 & 0.95 \\
							\midrule
							$L=4$ & 2 & 3.00 & 3.00 & 2.80 & 2.70 & 2.50 & 0.95 & 0.93 & 0.85 & 0.82 & 0.76 \\
							& 10 & 3.00 & 3.00 & 3.00 & 3.00 & 3.00 & 0.95 & 0.95 & 0.96 & 0.95 & 0.95 \\
							& 20 & 3.00 & 3.00 & 3.00 & 3.00 & 3.00 & 0.95 & 0.95 & 0.96 & 0.96 & 0.95 \\
							& 30 & 3.00 & 3.00 & 3.00 & 3.00 & 3.00 & 0.95 & 0.96 & 0.96 & 0.95 & 0.95 \\
							\midrule
							$L=5$ & 2 & 3.00 & 2.90 & 2.70 & 2.70 & 2.70 & 0.95 & 0.92 & 0.81 & 0.81 & 0.81 \\
							& 10 & 3.00 & 3.00 & 3.00 & 3.00 & 3.00 & 0.95 & 0.96 & 0.95 & 0.95 & 0.95 \\
							& 20 & 3.00 & 3.00 & 3.00 & 3.00 & 3.00 & 0.95 & 0.96 & 0.96 & 0.95 & 0.96 \\
							& 30 & 3.00 & 3.00 & 3.00 & 3.00 & 3.00 & 0.95 & 0.96 & 0.96 & 0.96 & 0.95 \\
							\bottomrule
						\end{tabular*}
					\end{table}
					
					\subsection{Low back pain data}\label{sec:low-back-pain}

					\begin{figure}[t!]
						\centering
						\includegraphics[width=0.6\textwidth, trim = 0 25 0 0, clip]{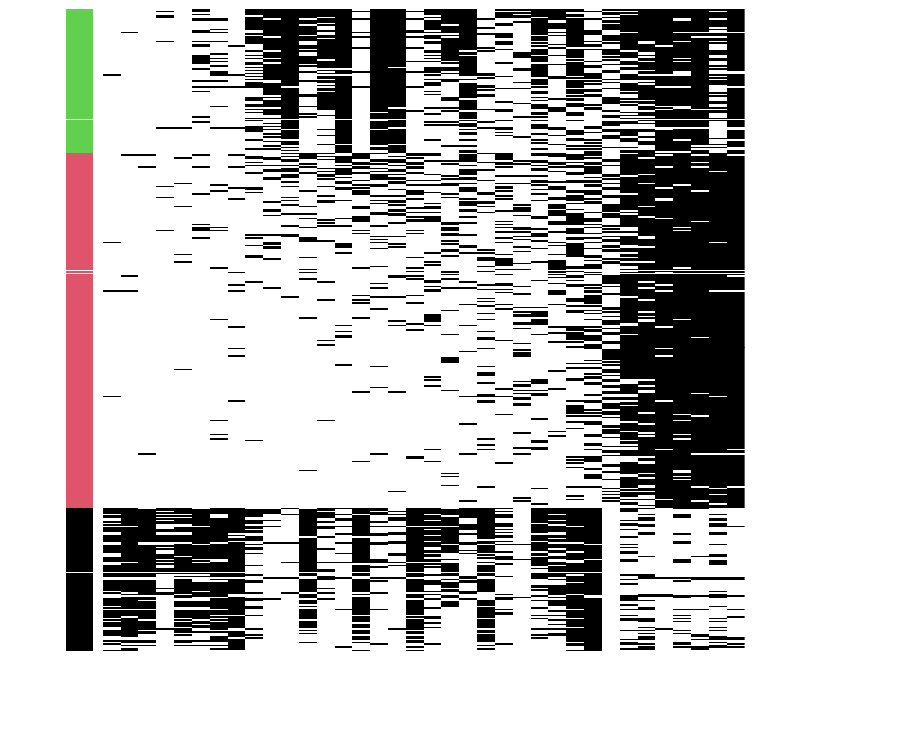}
						\caption{Low back pain data. Visualization of the original data
							matrix with patients in the rows and the binary variables in the
							columns. Patients are ordered by the expert-based grouping with
							the grouping indicated by color on the left and within-grouping by
							the average agreement level; columns are also ordered by the
							average agreement level. }\label{plot:back}
					\end{figure}
					
					\citet{lca:FopetAl2017} use model-based clustering to analyze a
					multivariate binary data set on low back pain for 425 patients on 36
					questions about the presence or absence of certain pain symptoms. An
					expert-based classification of the patients into three groups is
					available which serves as a gold standard. Figure~\ref{plot:back}
					provides a visualization of the binary data matrix, with the rows
					ordered by the expert-based grouping and within the grouping by
					average agreement level and columns ordered by average agreement
					level.

					\begin{table}[t!]
						\centering
						\caption{Low back pain data. Clustering results (number of clusters
							$K_+$ and ARI) for various specifications of $a_\mu$, $\cb$, and
							$L$.}\label{tab:back}
						\begin{tabular*}{\textwidth}{@{\extracolsep\fill}rr|rrrrr|rrrrr}
							\toprule
							\multicolumn{2}{c}{}& \multicolumn{5}{c}{$K_+$}&\multicolumn{5}{c}{ARI}\\
							\cmidrule{3-7}\cmidrule{8-12}%
							&&\multicolumn{5}{c}{$\cb$} &\multicolumn{5}{c}{$\cb$}\\
							&$a_\mu$	&1 & 10 & 20 & 30 & 40 & 1 & 10 & 20 & 30 & 40\\
							\midrule
							$L=2$& 2 & 11.00 & 6.00 & 5.00 & 5.00 & 4.00 & 0.37 & 0.46 & 0.49 & 0.65 & 0.80 \\
							& 10 & 9.00 & 5.00 & 4.00 & 3.00 & 3.00 & 0.38 & 0.45 & 0.68 & 0.82 & 0.80 \\
							& 20 & 9.00 & 5.00 & 4.00 & 3.00 & 3.00 & 0.37 & 0.53 & 0.68 & 0.80 & 0.79 \\
							& 30 & 9.00 & 5.00 & 3.00 & 3.00 & 3.00 & 0.39 & 0.43 & 0.82 & 0.79 & 0.80 \\
							\midrule
							$L=3$ & 2 & 8.00 & 5.00 & 4.00 & 4.00 & 4.00 & 0.45 & 0.49 & 0.79 & 0.79 & 0.80 \\
							& 10 & 8.00 & 3.00 & 3.00 & 3.00 & 3.00 & 0.36 & 0.74 & 0.81 & 0.81 & 0.82 \\
							& 20 & 9.00 & 5.00 & 3.00 & 3.00 & 3.00 & 0.53 & 0.40 & 0.81 & 0.72 & 0.80 \\
							& 30 & 8.00 & 4.00 & 3.00 & 3.00 & 3.00 & 0.34 & 0.59 & 0.81 & 0.73 & 0.79 \\
							\midrule	
							$L=4$ & 2 & 8.00 & 5.00 & 4.00 & 5.00 & 4.00 & 0.43 & 0.51 & 0.75 & 0.53 & 0.80 \\
							& 10 & 7.00 & 4.00 & 3.00 & 3.00 & 3.00 & 0.33 & 0.68 & 0.69 & 0.83 & 0.83 \\
							& 20 & 8.00 & 4.00 & 3.00 & 3.00 & 3.00 & 0.34 & 0.50 & 0.81 & 0.74 & 0.79 \\
							& 30 & 8.00 & 4.00 & 3.00 & 3.00 & 3.00 & 0.35 & 0.49 & 0.72 & 0.73 & 0.79 \\
							\midrule
							$L=5$ & 2 & 7.00 & 5.00 & 5.00 & 4.00 & 4.00 & 0.46 & 0.43 & 0.49 & 0.80 & 0.80 \\
							& 10 & 6.00 & 4.00 & 3.00 & 3.00 & 3.00 & 0.40 & 0.45 & 0.71 & 0.82 & 0.83 \\
							& 20 & 7.00 & 4.00 & 3.00 & 3.00 & 3.00 & 0.39 & 0.47 & 0.83 & 0.80 & 0.80 \\
							& 30 & 7.00 & 4.00 & 3.00 & 3.00 & 3.00 & 0.41 & 0.69 & 0.83 & 0.75 & 0.81 \\
							\bottomrule
						\end{tabular*}
					\end{table}
					
					For this data set, the presence
					of  within-cluster correlation is indicated by the following
					results. On the one hand, a standard LCA model fitted to this data set
					including all variables overestimates the number of clusters in
					the data, e.g., estimating the LCA model with maximum likelihood using
					the \textsf{R} package \pkg{poLCA} results in five components. On the
					other hand, fitting a standard LCA model to each expert-based class
					using maximum likelihood estimation, results in LCA models with two
					components for expert-based class 1 or 2, and one component for
					expert-based class~3.

					\citet{lca:FopetAl2017} pursue a variable selection approach to obtain
					an LCA model where classes correspond to the grouping induced by the
					expert-based classification (see Table~\ref{tab:backOthers}). Using
					the Bayesian mixture of LCA models approach, we also aim at
					identifying the expert-based grouping but without eliminating
					redundant variables which induce within-cluster correlation.

					The clustering results of the Bayesian mixture of LCA models approach
					are reported in Table~\ref{tab:back} for varying number of
					subcomponents $L$ and prior parameters $a_\mu$ and $\cb$. For each
					parameter combination, ten runs of the MCMC sampling scheme are
					performed and results are reported for the run selected using the
					procedure described in Section~\ref{sec:sim}.  The results are in line
					with those obtained for the simulation study with correlated
					clusters. Specifying shrinkage priors with $\cb\ge 20$, $a_\mu\ge 10$
					and $L \ge 3$, results in a good clustering performance with three
					clusters being estimated. 
						The ARI is at least 0.69 and the largest
						value obtained is 0.83. The largest value achieved 
					outperforms even
					the value of 0.79 reported by \cite{lca:FopetAl2017}. The alternative
					methods proposed for clustering multivariate categorical data --
					except for the variable selection approach proposed by
					\citet{lca:FopetAl2017} -- estimate between 5 and 10 clusters with an
					ARI around 0.50, see Table~\ref{tab:backOthers}. Thus, the Bayesian
					mixture of LCA models using priors inducing considerable amount of
					shrinkage succeeds in capturing the associated clusters in the back
					pain data set, avoiding the appearance of spurious clusters, which are
					included when fitting a standard LCA model due to the independence
					assumption.

					\begin{table}[t!]
						\centering
						\caption{Low back pain data. Alternative methods, with their default
							settings, 10 runs.}\label{tab:backOthers}
						\begin{tabular}{rrrr}
							\toprule
							& $K_+$ & ARI & err \\
							\midrule
							LCAvarsel & 3 & 0.73 & 0.09 \\
							BayesLCA & 5 & 0.50 & 0.39 \\
							PoLCA & 5 & 0.48 & 0.39 \\
							Rmixmod & 5 & 0.50 & 0.38 \\
							PReMiuM & 8 & 0.40  &  0.47 \\
							PReMiuM-contVS & 8 & 0.42 & 0.46 \\
							PReMiuM-binVS & 9 & 0.42 & 0.43 \\	
							\bottomrule
						\end{tabular}
					\end{table}
					
					\section{Summary}\label{sec:summary}
					
					In the Bayesian framework of model-based clustering of multivariate
					categorical data, this paper addresses the issue of dealing with
					associated variables within a cluster.  We propose to model each
					cluster by a LCA model capturing possible within-cluster associations.
					The resulting two-layer mixture of LCA models is completed through the
					specification of carefully selected shrinkage priors.
					The use of shrinkage priors supports the modeling aims in model-based
					clustering, i.e., they enable the identification of the two-layer
					model and the automatic estimation of a sparse number of clusters.
					Note that the proposed approach relies crucially on the specification
					of shrinkage priors. However, the defined shrinkage priors serve
					different purposes and can be thought of as a kind of regularization
					or soft constraint imposed on the different levels in the model
					estimation.
					
					Specifically, on the lower
					level shrinkage of the class occurrence probabilities forming
					one cluster towards the cluster center probabilities  allows identification of the cluster distributions. The
					likelihood of a two-layer mixture model is invariant to changing the
					assignment of subcomponents to components.  Shrinkage priors ensure
					that subcomponents are pulled towards a common cluster centroid, in
					this way inducing identifiability and destroying the invariance.
					On the upper level, shrinking the centers of the cluster distributions
					toward the uniform distribution supports the estimation of a small
					number of clusters, as the cluster distributions are enforced to be
					more similar. This shrinkage also acts as a kind of variable selection
					procedure; the occurrence probabilities of a single variable are
					thereby pulled towards the uniform distribution, reducing their
					contribution to the likelihood and suppressing in this way the
					appearance of additional noisy clusters.
					
					Finally, controlling against overfitting the number of components $K$
					is achieved by specifying appropriate priors on $K$ and the mixture
					weights which a priori will induce a large weight on partitions with a
					small number of clusters. The shrinkage prior for $K$ favors the
					homogeneity model and only allows a deviation from the one-cluster
					model if the heterogeneity in the data is strong enough.
					Additionally, the shrinkage prior on the mixture weights on the upper
					level obtained through the dynamic specification prevents that the
					number of clusters increases as fast as the number of proposed
					components.
					
					The telescoping sampler, where sampling the number of components $K$
					is conditionally independent of the component-specific parameters, has
					clear advantages when fitting a mixture distribution with complex
					component densities to data with an unknown number of clusters. The
					mixture of LCA models is a nice example for the application of the
					telescoping sampler as for this model it might be too cumbersome to
					design acceptable proposals for employing RJMCMC or integrate out the
					parameters for sampling the indicators marginally as required for the
					samplers used in Bayesian non-parametric
					mixture analysis.
					
					Extensions of the proposed approach could be considered in several
					directions.  The clustering of multivariate categorical data often
					involves many variables. Co-clustering \citep{lca:Govaert+Nadif:2013},
					also known as biclustering or two-mode clustering, aims at
					simultaneously grouping rows and columns of a rectangular data
					matrix. Simultaneously clustering rows and columns of a multivariate
					categorical data matrix eases interpretation by providing a natural
					grouping of variables. It would thus be of interest to extend the
					proposed Bayesian mixture of LCA models approach with a prior on the
					number of components to also cluster the columns. Alternatively one
					could also investigate a Bayesian two-layer mixture model for modeling
					multivariate count data, e.g., using mixtures of independent Poisson
					distributions on the lower level.

					\begin{appendices}
						\section{The beta-negative-binomial distribution}\label{sec:beta-negat-binom}
						
						The beta-negative-binomial (BNB) distribution is a three-parameter
						distribution with support on the non-negative integers.  The
						distribution results as a hierarchical generalization of the Poisson,
						the geometric and the negative-binomial distribution.
						
\begin{figure}[t!]
							\centering
							\includegraphics[width=0.98\textwidth]{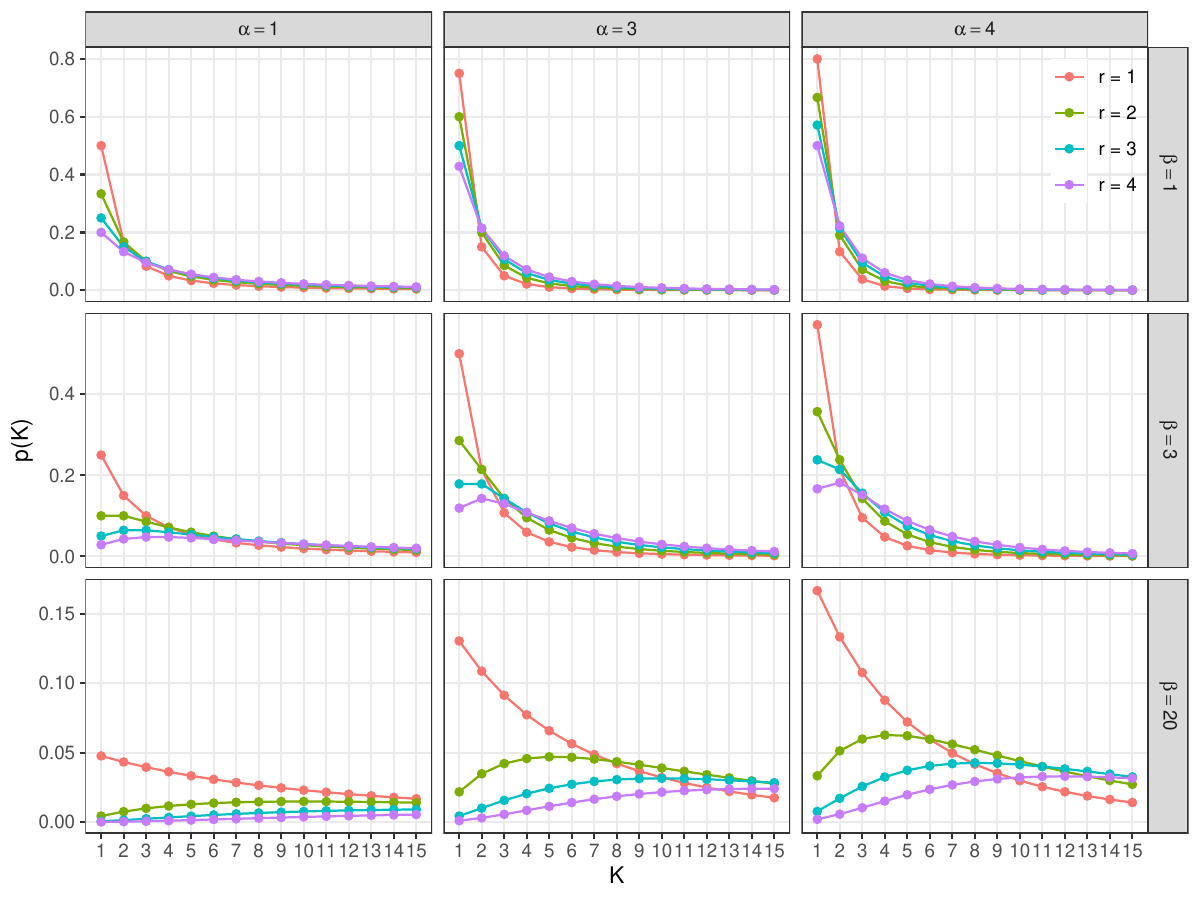}
							\caption{Priors $p(K)$ derived from the translated prior
								$K-1 \sim \mathit{BNB}(r, \alpha, \beta)$ for
								various parameter values.}\label{plot:BNB}
						\end{figure}

						A translated version is considered for the number of components $K$
						with the probability mass function corresponding to
						$K-1 \sim \mathit{BNB}(r, \alpha, \beta)$ given by:
						\begin{align}
							p(K) &=  
							\frac{\Gamfun{r + K -1}\text{B}(r+\alpha, K -1 + \beta)}{ \Gamfun{r} \Gamfun{K} \text{B}(\alpha,\beta)}  .
							\label{BNB}
						\end{align}
						The mean of $K$ exists for $\alpha > 1$ and is equal to
						$1 + r \frac{\beta}{\alpha - 1}$. The different shapes of the BNB
						distribution obtained for various values of the parameters are
						illustrated in Figure~\ref{plot:BNB}.
						
						\section{MCMC sampling scheme}\label{sec:mcmc-scheme}
						
						Posterior inference for the Bayesian mixture of LCA models is
						performed through MCMC sampling based on data augmentation and Gibbs
						sampling. 
						First, the number of components $K$ is considered a latent variable. Second, to
						indicate the component to which each observation belongs,
						latent allocation variables $\bS=(S_1,\ldots,S_N)$ taking values in
						$\{1,\ldots,K\}$ are introduced such that
						\begin{align*}
							p(\by_i|\btheta_1,\ldots,\btheta_K,S_i=k)&=p_k(\by_i|\btheta_k).
						\end{align*}
						Additionally, latent allocation variables $\bI = (I_1,\ldots,I_N)$
						taking values in $\{1,\ldots,L\}$ are introduced to indicate the
						subcomponent to which an observation within a component is
						assigned. This implies
						\begin{align*}
							p(\by_i|S_i = k, \btheta_k, I_i = l) &= \prod_{j=1}^r
							\prod_{d=1}^{D_j} \pi_{kl,jd}^{I\{y_{ij} = d\}}.
						\end{align*}
						The latent variables $(\bS, \bI)$ and parameters
						$(K,\bm{\eta}_K, \alpha, \bm{w}_{KL}, \bm{\pi}_{KL},
						\bm{\phi}_K,\bm{\mu}_K, \bm{b}_{\phi})$ are sampled using the following
						Gibbs sampling scheme after suitable initialization.
						
						The MCMC scheme is initialized by assigning observations to a
						component as well as to subcomponents within a component, i.e., by
						specifying the latent variables $(\bS, \bI)$.  The assignment vector
						$\bS$ is determined using $k$-means clustering starting
						with an overfitting
						number of clusters (e.g., $K = 10$ in the empirical
						demonstrations). Within each component, the subcomponent assignment
						vector $\bI$ is then obtained using random assignment with equal
						probabilities to initially obtain comparable subcomponent
						distributions within a component.  Based on the classification induced
						by $(\bS, \bI)$, the initial occurrence probabilities are obtained as
						the empirical occurrence probabilities within each of the
						subcomponents. The initial values for $\bmu_{k,j}$ and $\phi_{k,j}$
						are set to the vector of length $D_j$ with identical values $1/D_j$
						and $D_j$, respectively.
						
						\vspace*{2mm}
						
						\noindent The MCMC sampling scheme then proceeds as follows:
						\begin{enumerate}	
							\item For a given $K$, update the partition of the data:
							\begin{enumerate}
								\item Sample $S_i$ for $i=1,\ldots,N$ from
								$p(S_i |K, \boldeta_K,\btheta_1,\ldots, \btheta_K,\ym)$,
								marginalized w.r.t. the subcomponent indicator $\bI$,
								using:
								\[	\Prob{S_i=k|K, \etav_K,\btheta_1,\ldots, \btheta_K, \ym_i}  \propto \eta_k p_k(\ym_i|\btheta_k), \quad k=1,\ldots,K,\]
								where  $p_k(\by_i | \btheta_k)$  is  the cluster density given by a LCA model:
								\[ 	p_k(\by_i| \btheta_k) = \sum_{l=1}^{L} w_{kl} \prod_{j=1}^r \prod_{d=1}^{D_j} \pi_{kl,jd}^{I\{y_{ij}=d\}}.
								\]
								
								\item Determine $K_+ = \sum_{k=1}^K I\{N_k>0\}$, where  $N_k=\#\{i|S_i=k\}$.
								\item Relabel the components such that the first $K_+$ components are
								non-empty.
							\end{enumerate}
							
							\item Conditional on $\bS$, update the parameters and hyperparameters
							for each filled cluster $k=1,\ldots,K_+$:
							\begin{enumerate}
								\item[(a)] Sample subcomponent indicators
								from
								\begin{align*}
									\Prob{I_i=l|S_i=k, \by_i, \cdot 
									} &\propto w_{kl}
									\prod_{j=1}^r \prod_{d=1}^{D_j} \pi_{kl,jd}^{I\{y_{ij}=d\}}, \quad l=1,\ldots,L.
								\end{align*}
								\item[(b)]  Sample subcomponent weights from
								\begin{align*}
									\bm{w}_k|\bI, \bS \sim \mathcal{D}_{L}(d_{k1},\ldots,d_{kL}), \quad l=1,\ldots,L,
								\end{align*}
								using $d_{kl}=\delta+N_{kl}$ where $N_{kl}=\#\{i | I_i=l, S_i=k\}$ is
								the number of observations allocated to subcomponent $l$ in cluster
								$k$.
								\item[(c)] Sample subcomponent occurrence probabilities from
								\[ \bm{\pi}_{kl,j}|\bmu_{k,j},\phi_{k,j},\bS,\by \sim \mathcal{D}_{D_j}(\ba_{kl,j}), 	\]
								where
								\begin{align*}
									\ba_{kl,j}&=\bmu_{k,j}\phi_{k,j}+a_{00}\bm{1}_{D_j}+\bm{N}_{kl,j},
								\end{align*}
								with $\bm{N}_{kl,j}=(N_{kl,j1},\ldots, N_{kl,jD_j})$ and where
								$N_{kl,jd}$ indicates how often in subcomponent $l$ of cluster $k$
								category $d$ is observed for feature $j$:
								\begin{align*}
									N_{kl,jd}&=\sum_{\iota \in \{i|S_i=k, I_i=l\}} I\{y_{\iota j}=d\}.
								\end{align*}
								\item[(d)] Sample hyperparameters for $j=1,\ldots,r$:
								\begin{itemize}
									\item[(-)] Sample $\bmu_{k,j}$ from
									$p(\bmu_{k,j}|a_\mu)\cdot \prod_{l=1}^L
									p(
									\bm{\pi}_{kl,j}   |\bmu_{k,j},\phi_{k,j},a_{00})$, i.e.,
									\begin{align*}
										&\propto \mu_{k,j1}^{a_\mu-1}\cdots \mu_{k,jD_j}^{a_\mu-1} \prod_{l=1}^L \mathcal{D}_{D_j}(\bm{\pi}_{kl,j}|\bmu_{k,j} \phi_{k,j}+a_{00}\bm{1}_{D_j})
									\end{align*}
									using a Metropolis-Hastings step with Dirichlet proposal density
									$$\bmu_{k,j}^{new} \sim \mathcal{D}_{D_j}(\bmu_{k,j}^{old}s_\mu),$$
									where $s_\mu$ is a fixed precision value acting as a calibration
									parameter.
									\item[(-)]
									Sample $\phi_{k,j}$ from $p(\phi_{k,j}|a_\phi,b_{\phi_j}) \cdot  \prod_{l=1}^L p
									(\bm{\pi}_{kl,j}|\bmu_{k,j},\phi_{k,j},a_{00})$,\\
									i.e.
									\[	\propto \phi_{k,j}^{-a_\phi-1} \exp(- b_{\phi_j}/ \phi_{k,j}) \cdot \prod_{l=1}^L \mathcal{D}_{D_j}(\bm{\pi}_{kl,j}|\bmu_{k,j} \phi_{k,j}+a_{00}\bm{1}_{D_j})\]
									using a random walk Metropolis-Hastings step with proposal
									$$ \log(\phi_{k,j}^{new}) \sim \cN(\log(\phi_{k,j}^{old}),s^2_\phi), $$
									where $s^2_\phi$ is a fixed variance value acting as a calibration
									parameter.
								\end{itemize}
							\end{enumerate}
							
							\item Sample the dimension-specific hyperparameter for the precision
							of the component location for $j=1,\ldots,r$:
							\begin{itemize}
								\item[(-)] Sample $b_{\phi_j}$ from
								\begin{align*}
									b_{\phi_j}  &\sim \cG(c_{\phi_j},d_{\phi_j}),
								\end{align*}
								where
								\begin{align*}
									c_{\phi_j} &= \cb + K_+ a_\phi,&
									d_{\phi_j} &= \db + \sum_{k=1}^{K_+} \frac{1}{\phi_{k,j}}.
								\end{align*}
							\end{itemize}
							
							\item Conditional on $\cC$,  sample   $K$ and $\alpha$:	
							\begin{enumerate}
								\item Sample $K$ from
								\begin{align*} 
									p(K| \cC, \alpha)  &\propto  p(K)
									\frac{ \alpha ^{K_+}   K!}{K ^{K_+} (K-K_+)!} \prod_{k=1}^{K_+} \frac{\Gamma(N_k+\frac{\alpha}{K})}{\Gamma(1+ \frac{\alpha}{K})}	,
									\quad K=K_+, K_+ +1,\ldots,K_{\max}.
								\end{align*}
								In theory, $K_{\max}$ is infinite. For the computational
								implementation, a finite value is pre-specified. A value large
								enough to capture all relevant values of $K$ needs to be selected,
								but as small as possible to reduce computational complexity.
								\item Use a random walk Metropolis-Hastings step with proposal
								$\log(\alpha^{new}) \sim \cN({\log(\alpha^{old}),s_\alpha^2})$ to
								sample $\alpha|\cC,K$ from
								\begin{align*}
									p(\alpha|\cC,K) &\propto p(\alpha)\frac{ \alpha ^{K_+} \Gamma(\alpha)}{\Gamma(N+\alpha)}
									\prod_{k=1}^{K_+} \frac{\Gamma(N_k+\frac{\alpha}{K})}{\Gamma(1+ \frac{\alpha}{K})}.
								\end{align*}
							\end{enumerate}
							\item Add $K-K_+$ empty components and update the component weight
							distribution conditional on $K, \cC, \alpha$ and $\phi$:
							\begin{enumerate}
								\item If $K> K_+$, then add $K-K_+$ empty components (i.e., $N_k=0$
								for $k=K_+ +1,\ldots,K$) by sampling $\bm{w}_k$ from the 
								symmetric Dirichlet prior for
								$k=K_+ +1,\ldots,K$:
								\begin{align*}
									\bm{w}_k|d_0 \sim 
									\mathcal{D}_{L}(\delta),
								\end{align*}
								and sampling $\btheta_k$ for $k=K_+ +1,\ldots,K$ from the
								hierarchical prior defined for
								$\bm{\pi}_{kl,j}| b_{\phi_j}, a_\phi, a_\mu, a_{00}$ for
								$j=1,\dots,r$, conditional on the current value of $b_{\phi_j}$:
								\begin{align*}
									\bmu_{k,j}|a_\mu &\sim 
									\mathcal{D}_{D_j}(a_\mu),\\
									\phi_{k,j}|a_\phi,b_{\phi_j} &\sim \cG^{-1}(a_\phi,b_{\phi_j}),\\
									\bm{\pi}_{kl,j}|a_{00}, \bmu_{k,j}, \phi_{k,j} &\sim 
									\mathcal{D}_{D_j}(\bmu_{k,j}\phi_{k,j}+a_{00}\bm{1}_{D_j}), \quad l=1,\ldots,L.
								\end{align*}
								\item Sample
								\begin{align*}
									\boldeta_K|K,\alpha,\cC &\sim \mathcal{D}_K(\gamma_1,\ldots,\gamma_K),
								\end{align*}
								where $\gamma_k=\alpha/K + N_k $.
							\end{enumerate}
						\end{enumerate}
					\end{appendices}

\section{References}
\bibliographystyle{Chicago}
\bibliography{lca}

\end{document}